\documentclass[conference,compsoc]{IEEEtran}
\usepackage{algorithm}
\usepackage{algpseudocode}
\usepackage{algorithmicx}
\usepackage{amsmath}
\usepackage{amsthm}
\usepackage{amssymb}
\usepackage{mathtools}
\usepackage{bbm}
\usepackage{subfigure}

\usepackage{ulem}

\newtheorem{thm}{Theorem}
\newtheorem{defi}{Definition}
\newtheorem{prop}{Proposition}

\newtheorem{rmk}{Remark}

\newcommand{\wz}[1]{{\textcolor{blue}{[wz: #1]}}}

\usepackage{tikz}
\usepackage{amsmath}

\DeclareMathOperator*{\argmax}{argmax}

\ifCLASSOPTIONcompsoc
  \usepackage[nocompress]{cite}
\else
  \usepackage{cite}
\fi
\ifCLASSINFOpdf
\else
\fi
\begin{document}
%
\title{Meeting Utility Constraints in Differential Privacy: A Privacy-Boosting Approach
}

\author{\IEEEauthorblockN{Bo Jiang\IEEEauthorrefmark{1}\IEEEauthorrefmark{2}, Wanrong Zhang\IEEEauthorrefmark{1}\IEEEauthorrefmark{2},
Donghang Lu\IEEEauthorrefmark{1}, Jian Du\IEEEauthorrefmark{1}, 
Sagar Sharma\IEEEauthorrefmark{1}, and
    Qiang Yan\IEEEauthorrefmark{1}.}
\IEEEauthorblockA{\IEEEauthorrefmark{1}
TikTok Inc, \IEEEauthorrefmark{2} Equal Contribution.\\ Email: \{bojiang, wanrongzhang, Donghang.lu, jian.du, sagar.sharma, yanqiang.mr\}@tiktok.com}}

\maketitle

\begin{abstract}
Data engineering often requires accuracy (utility) constraints on results, posing significant challenges in designing differentially private (DP) mechanisms, particularly under stringent privacy parameter $\epsilon$. In this paper, we propose a privacy-boosting framework that is compatible with most noise-adding DP mechanisms. Our framework enhances the likelihood of outputs falling within a preferred subset of the support to meet utility requirements while enlarging the overall variance to reduce privacy leakage. We characterize the privacy loss distribution of our framework and present the privacy profile formulation for $(\epsilon,\delta)$-DP and Rényi DP (RDP) guarantees. We study special cases involving data-dependent and data-independent utility formulations. Through extensive experiments, we demonstrate that our framework achieves lower privacy loss than standard DP mechanisms under utility constraints. Notably, our approach is particularly effective in reducing privacy loss with large query sensitivity relative to the true answer, offering a more practical and flexible approach to designing differentially private mechanisms that meet specific utility constraints.
\end{abstract}

\IEEEpeerreviewmaketitle

\section{Introduction}

Differential Privacy (DP) \cite{Dwork2006} has emerged as the leading-approach in privacy-preserving data analysis. It offers robust privacy guarantees of individual data points even in the presence of adversaries with significant auxiliary information. A typical DP mechanism achieves this by adding noise to the true answers of queries, ensuring that the output is sufficiently obfuscated. The magnitude of noise required to meet a specified privacy level is typically determined by the privacy parameter $\epsilon$: higher privacy levels, corresponding to small $\epsilon$, require more noise. This often comes with a significant trade-off: the added noise can substantially degrade the utility of the data even to the point where the results become impractical or even useless for analysis. 
Traditional DP mechanisms determine the noise magnitude based on a target privacy parameter and then provide corresponding utility guarantees. This method, however, might fail to align with the practical needs of analysts who have specific utility constraints they aim to satisfy.

Our research is motivated by the practical considerations of deploying DP in real-world scenarios. Analysts usually have target utility constraints according to the specific applications, and hope that DP mechanisms can satisfy these constraints as closely as possible. Often, these utility formulations are often more complex than simply considering ``the variance of the noise" or ``absolute distance." Although absolute error can easily be converted into other error formats, the tolerance for absolute error is dependent on the true values. For instance, when relative error is of greater concern, the acceptable level of absolute error will vary accordingly.
Consider a mobile health application that collects user data for tracking physical activity levels to provide personalized health insights. To protect user privacy, a DP mechanism is applied to the collected data. A health regulator may mandate that errors introduced by DP mechanisms in reporting user activity should not exceed 5\% of the true value to ensure the reliability of health insights.

Conventional approaches primarily focus on achieving privacy guarantees without regard of utility constraints, which we refer to as a privacy-first DP approach. This method designs the noise distribution according to given privacy parameters and then calculates the corresponding utility with the noise distribution. Although there are several utility-first approaches \cite{ligett2017accuracy,whitehouse2022brownian} that conversely searches for the minimal privacy parameter to achieve the desired utility, they typically consider fixed forms of noise distribution, which may not be optimal for specific utility requirements. Therefore, there is a pressing need for DP mechanisms that can dynamically balance privacy and utility based on the specific requirements of the analysis.

In this paper, we introduce a novel framework that incorporates a target utility constraint, instead of adhering to a fixed noise magnitude determined solely by the privacy parameter. To formalize this, we introduce a mechanism that ensures outputs fall within preferred regions tailored to the true query answers. This preferred region is denoted as $\mathcal{S}(Q(X))$, which is co-determined by the form of region $\mathcal{S}$, query $Q$, and dataset $X$.
Our goal is to design a differentially private mechanism $\mathcal{M}_{pb}$ ($pb$  stands for privacy boosting) such that for every dataset $X \in \mathcal{X}$, we have the following utility guarantee for the query answer $Q(X)$: 
\begin{equation}\label{eq:utility}
    \text{Pr}[\mathcal{M}_{pb}(X) \in \mathcal{S}(Q(X)] \ge \rho,
\end{equation}
where $\rho$ is a confidence level indicating the likelihood that the noisy output falls within in $\mathcal{S}(Q(X)$. Concurrently, we hope $\mathcal{M}_{pb}$ incurs reduced privacy loss compared to standard DP noise-adding mechanisms that achieves this. This formulation ensures that the Privacy-Boosting differentially private (PB-DP) mechanism produces outputs that are not only private but also practically useful. 

Our mechanism consists of a kernel DP mechanism that can instantiate any standard noise-adding mechanism. Given an input, it simply boosts the probability density function (pdf) within the preferred region with a boosting rate $q$. This $q$ can be determined by the kernel mechanism and the target utility constraint. Thus, the resulting noise distribution is a re-weighted distribution of that in the kernel mechanism, enhancing the likelihood of outputs falling within the preferred region. That being said, for a given utility constraints, there are multiple combinations of the kernel DP mechanism and the correspond $q$. Our mechanism then searches for the optimal pair of them that yields the minimal privacy loss.

The privacy analysis of our algorithm is critical. We characterize the privacy loss distribution (PLD) of our mechanism as a function of the kernel DP mechanism and the boosting rate $q$. From there, we demonstrate that our mechanism satisfies differential privacy. We further provide the formulation of the privacy profile for 
$(\epsilon,\delta)$-DP and RDP guarantee. 


Our results show that under the target utility constraint, our mechanism can significantly reduce the overall privacy loss compared  to directly applying its kernel DP mechanism. The underlying reason is that our mechanism adopts a much smaller privacy parameter for the kernel mechanism, and achieves the utility constraints by strategically boosting the probability density within the preferred region. Intuitively, our mechanism enlarges the overall noise variance to reduce privacy leakage, and concurrently boosting the likelihood of preferred region to meet the utility constraints.

To illustrate our approach, we explore three special cases of preferred regions.

\noindent{\textbf{Data-dependent preferred region}} The preferred region $\mathcal{S}(Q(X))$ varies depending on the true value $Q(X)$. An application of this case is bounded relative error, where a smaller range of $\mathcal{S}(Q(X))$ is preferred when $Q(X)$ is small. Only a few works have focused on this problem due to the inherent difficulty of making the noise distribution dependent on the true answer without violating privacy. Notably, iReduct\cite{10.1145/1989323.1989348} attempted to address this by iteratively adjusting the noise magnitude based on query sensitivity and data distribution. However, the mechanism features high computation cost, and the boosted utility is without guarantee. In contrast, our mechanism features efficient one-shot release sampled from a fixed distribution,  meeting a pre-defined utility constraint.

\noindent{\textbf{Data-independent preferred region}} The preferred region $\mathcal{S}(Q(X))$ is independent of the true value $Q(X)$. Bounded noise mechanisms, such as the bounded Laplacian mechanism proposed by Geng et al. \cite{Geng2018TruncatedLM}, address this problem by bounding the noise magnitude. However, these mechanisms often requires a large failure probability to handle the output support discrepancies for neighboring datasets, a challenge that is even amplified under composition. Our mechanism, on the other hand, provides statistically soft boundaries that help align the supports of neighboring datasets, and therefore features a much desirable privacy profile.

\noindent{\textbf{Deterministic preferred region}} The preferred region $\mathcal{S}(Q(X))$ is deterministic for any possible $X\in\mathcal{X}$. For example, valid outputs might have a bounded support. A classical way to solve this problem is by bounded support or truncated mechanisms, that involves  truncation in post-processing and resampling during release. These mechanisms been widely studied in literature\cite{Geng2018TruncatedLM, 2022-2-7-0193, holohan2018bounded}. However, truncated mechanism leads to a large likelihood to release an output at the boundaries, providing limited utility. The design of bounded support mechanisms with resampling must be considered case-by-case according to the specific distributions. At the same time, the increased likelihood for values in the bounded support also enlarges the overall privacy leakage. Conversely, our framework features a general design for most of the additive noise distribution. Also, the enlongated tail in the noise distribution reduces the overall leakage.
In summary, our contributions are fourfold:
\begin{enumerate}
    \item We introduce a framework for designing differentially private mechanisms that prioritize utility that might be a data-dependent measure. The kernel mechanism in our framework is versatile with most of the noise-adding mechanisms, such as the Gaussian and Laplace mechanisms.
    \item We provide a detailed privacy analysis including characterizations of the PLD, and privacy profile formulations for $(\epsilon,\delta)$-DP and RDP guarantees, and further study the composability of our framework. We theoretically show that our framework achieves smaller privacy loss compared to only using its kernel mechanism under the same utility constraints.
    \item We explore three special cases corresponding to different specifications of the preferred output regions, including data-dependent, data-independent, and fixed preferred regions. Further, we show the potential of applying our framework in a local setting for frequency estimation, to provide accurate estimation both data value and category collection. These demonstrate the effectiveness of our mechanism in various scenarios.
    \item We demonstrate through extensive experiments that our mechanism can achieve lower privacy loss compared to standard mechanisms under given utility constraints, offering a more practical and flexible approach to differential privacy. 
\end{enumerate}

\subsection{Related works}

In recent years, Several research efforts have focused on optimizing noise-adding mechanisms to enhance the trade-off between privacy and utility, each approaching the challenge from different angles and perspectives.


Awan and Vadhan \cite{awan2023canonical} introduced the concept of canonical noise distribution (CND). This framework constructs one-dimensional additive noise mechanisms satisfying $f$-DP \cite{dong2019gaussian}, without wasting any privacy budget. A key benefit of considering $f$-DP is its lossless composition guarantees. Subsequent research \cite{awan2022log} delves into constructing log-concave canonical noise distributions, as log-concave ensures that higher outputs of the mechanism correspond to higher input values. Notably, their CND for $(\epsilon,0)$-DP aligns with one of the staircase mechanisms proposed in prior works \cite{geng2015optimal,geng2015staircase}.

The design of optimal noise-adding distributions, particularly staircase-shaped densities for $(\epsilon,0)$-DP, has been explored to minimize the worst-case query cost across all possible outputs\cite{geng2015optimal,geng2015staircase,soria2013optimal}. While such frameworks aim to minimize worst-case query costs across all possible outputs, their applicability to practical scenarios is limited, as they may not always align with specific utility requirements. In contrast, our approach prioritizes boosting the likelihood of noisy answers falling into preferred regions.

Jiang et al. \cite{jiang2024budget} introduce a budget recycling mechanism geared towards enhancing the probability of obtaining valid noisy query response within an acceptable absolute error range. Our approach extends this concept and caters to a broader range of utility requirements including relative error, fixed output domain, discrete-value in local settings.

iReduct \cite{10.1145/1989323.1989348} 
proposed an iterative approach to reduce the relative error by adjusting amount of noise added to the query answer, employing standard DP noise-adding mechanisms like Laplace mechanism. This approach seeks to identify minimal noise scales while maintaining consistent noise distribution shapes. Our approach instead directly modifies the probability density function of noise distributions to meet specific utility requirements.

Additionally, iterative searching approaches have been employed to enhance the accuracy of private empirical risk minimization (ERM) algorithms \cite{ligett2017accuracy,whitehouse2022brownian}. This ``noise reduction'' framework aim to explore privacy levels to meet accuracy constraints using ex-post privacy, a weaker and data-dependent variant of differential privacy. In contrast, our approach adheres to the worst-case differential privacy guarantees.

Smooth sensitivity \cite{nissim2007smooth} provides a nuanced approach to quantifying query sensitivity, considering the specific structure and distribution of data. While traditional measures like global sensitivity may be overly conservative, smooth sensitivity provides a more refined sensitivity metric, aligning with our approach of adjusting noise distributions based on specific utility requirements and data characteristics. This adaptability allows us to provide more refined privacy-utility trade-offs tailored to the unique context of each dataset and analysis task.

\section{Preliminaries of Differential Privacy}
In this section, we will provide an overview of differential privacy and introduce key techniques that are essential for understanding our work. 


Differential privacy \cite{Dwork2006} is a mathematical notion of database privacy, which ensures that the presence or absence of a single database item does not significantly affect the outcome of any analysis, thereby protecting individual data entries. 

\begin{defi}[$(\epsilon,\delta)$-DP \cite{Dwork2006}]\label{def.dp}
	A randomized algorithm $\mathcal{M}: \mathcal{X} \rightarrow \mathcal{R}$ is $(\epsilon,\delta)$-differentially private if for every pair of datasets $X,X' \in \mathcal{X}$ that arbitrarily differ in the values at most one entry, and for every subset of possible outputs $S \subseteq \mathcal{R}$,
	$\Pr[\mathcal{M}(X) \in \mathcal{S}] \leq \exp(\epsilon)\Pr[\mathcal{M}(X') \in S] + \delta.$
\end{defi}

Here, the privacy is quantified by a privacy-loss parameter $\epsilon$, and $\delta\in[0,1]$ is a small additive slack term. When $\delta = 0$, we call it (pure) $\epsilon$-DP. When $\delta > 0$, we often refer to it approximate DP. Smaller $\epsilon$ and $\delta$ imply stronger privacy guarantees.

Differential privacy is usually achieved by carefully introducing randomness into the computation. A common class of differentially private mechanisms is noise-adding mechanism, for example, the Gaussian mechanism. 
The scale of the noise is determined by the sensitivity of the query. The sensitivity of a query $Q$ is defined as the maximum change in $Q$ between two neighboring datasets, which can only differ in one data entry: $\Delta Q = \max_{X,X':||X-X'||\le 1} | Q(X) - Q(X')|$ . The Gaussian mechanism with parameters $(\epsilon, \delta)$ takes in a function $Q$, dataset $X$, and outputs $Q(X)+\mathcal{N}(0,\sigma^2)$, where $\sigma=\sqrt{2\log(1.25/\delta)}\Delta Q/\epsilon$.

A fundamental property of differential privacy is composition, which studies how privacy guarantees degrade when multiple differentially private mechanisms are applied to the same dataset. When multiple analyses are performed, the total privacy loss accumulates. The basic composition theorem \cite{dwork2006our} suggests that $\epsilon$, $\delta$ both increase linearly with the number of queries, or sublinearly with advanced composition  \cite{dwork2010boosting, kairouz2015composition, murtagh2016complexity}. This large privacy loss consumption makes deploying DP solutions challenging in real-world applications, especially when the number of computation is large. Therefore, a tighter privacy accounting is desired as we can achieve better utility while still maintaining strong privacy guarantees when the cumulative privacy loss can be more accurately controlled. 

In recent years, other variants of differential privacy have been proposed to address various shortcomings of $(\epsilon, \delta)$-DP regarding composition. Some standard variants include R\'enyi DP (RDP) \cite{mironov2017renyi}, $f$-DP \cite{dong2019gaussian}, and zero-concentrated differential privacy (zCDP) \cite{dwork2016concentrated,bun2016concentrated}. We present the formal definition of RDP below. 

\begin{defi}[Rényi DP \cite{mironov2017renyi}]

A randomized mechanism $\mathcal{M}$ is $(\alpha,\epsilon)$-R\'enyi differentially private (RDP) for $\alpha>1$ if for all neighboring datasets $X$ and $X'$ :
    \begin{equation*}
        D_{\alpha}(P_{\mathcal{M}(X)}(y_n)||Q_{\mathcal{M}(X')}(y_n))\le{\epsilon},
    \end{equation*}
    where the Rényi divergence between two distribution $P$ and $Q$ is 
    \begin{equation*}
        D_{\alpha}(P||Q) = \frac{1}{1-\alpha} \mathbb{E}_{P}\left[\left(\frac{P}{Q}\right)^{\alpha-1}\right].
    \end{equation*}
\end{defi}

RDP offers analytical convenience for efficient privacy analysis, as the optimal composition for RDP is simply additive in $\epsilon$, for a fixed $\alpha$.

To analyze the privacy loss, an critical tool is the privacy loss distribution (PLD) introduced by Sommer et al.\cite{cryptoeprint:2018/820}, which is a probabilistic measure of privacy loss , and its framework provides a way to precisely quantify the cumulative privacy loss. 



\begin{defi}[Privacy Loss Distribution \cite{cryptoeprint:2018/820}]

The privacy loss random variable for a pair of neighboring datasets $X,X'$ under mechanism $\mathcal{M}$ is defined as $\Gamma_{X,X'} \overset{\Delta}{=} \log \frac{P_{\mathcal{M}(X)}(y)}{P_{\mathcal{M}(X')}(y)}$, where $y\sim \mathcal{M}(X)$. Here we use $\mathcal{M}(X)$ to denote the probability distribution of the mechanism's output. Similarly, we have $\Gamma_{X',X} \overset{\Delta}{=} \log \frac{P_{\mathcal{M}(X')}(y)}{P_{\mathcal{M}(X)}(y)}$, where $y\sim \mathcal{M}(X')$. The privacy loss distribution (PLD) is the distribution of $\Gamma_{X,X'}$, denoted $f_{\Gamma_{X,X'}}(\gamma)$. 
\end{defi}

Instead of considering the worst-case privacy loss for each mechanism, PLD allows for a more refined calculation of the total privacy loss by considering the distribution of privacy losses across all compositions. Given a PLD of a mechanism, we can convert it to the standard $(\epsilon, \delta)$-DP guarantees. The relationship is defined as the privacy profile \cite{DBLP:journals/corr/abs-1807-01647}.


\begin{defi}[Privacy profile \cite{DBLP:journals/corr/abs-1807-01647}] 
Given the privacy loss random variable for any pair of neighboring datasets $X,X'$: $\Gamma_{X/X'}$ and $\Gamma_{X'/X}$, let $\Gamma_{X,X'}\overset{\Delta}{=} \max\{\Gamma_{X/X'}, \Gamma_{X'/X} \}$.
The privacy profile is as follows: 
\begin{equation*}
\begin{aligned}
    \delta \ge & \max_{X,X'}\mathbb{E}_{\Gamma_{X,X'}}[\max\{0,1-\exp(\epsilon-\gamma)\}]\\
    =&\int_{\epsilon}^{\infty}(1-\exp(\epsilon-\gamma))f_{\Gamma_{X,X'}}(\gamma)d\gamma.\\
\end{aligned}
\end{equation*}

\end{defi}

PLD can also contribute to the accounting of DP variants, such as RDP. 
The RDP privacy parameter $\epsilon$ in terms of PLD $f_{\Gamma_{X,X'}}$ is given by
\begin{equation*}
    \epsilon = \frac{1}{1-\alpha} \max_{X,X'}\log\left(\int_{-\infty}^{\infty} e^{(\alpha - 1)\gamma}f_{\Gamma_{X,X'}}(\gamma) d\gamma\right).
\end{equation*}


In additional to these analytical approaches, numerical composition accounting are becoming increasingly popular \cite{cryptoeprint:2018/820}.
The core of this technique is based on the fact that the PLD of composed DP mechanisms is equivalent to the convolution of the PLD of those DP mechanisms. To accurately derive the convolution, Koskela et al. \cite{Koskela2019ComputingTD, Koskela2020TightAD} propose using FFT-based algorithms that treat the PLD as a time series signal and numerically calculate the cumulative leakage in the frequency domain. The latest work in this direction is by Zhu et al. \cite{Zhu2021OptimalAO}, who propose an analytic Fourier accounting algorithm deploying the characteristic function. This method overcomes a limitation of the FFT-based algorithm, which involves exhaustively searching for all neighboring datasets in the worst-case PLD scenario.

\section{Privacy Boosting Framework}

The utility of a DP mechanism is typically measured by the absolute error between the true value and the noisy output. Since most noise-adding mechanisms are mean-zero, the absolute error is typically solely captured by the variance of the noise. However, for many use cases, the absolute error without considering the true value does not provide enough meaningful indication on the accuracy: Low absolute error for small true value does not imply high accuracy, and high absolute error for large true value may not be unacceptably inaccurate. 
Therefore, we instead consider a data-dependent utility measure, formally given as follows. Let $Q(X)$ denote the true query answer. We have a specific preferred region associated with the true value, denoted as $\mathcal{S}(Q(X))$.

Our goal is to design a differentially private mechanism $\mathcal{M}$ such that for every dataset $X \in \mathcal{X}$, we have
\begin{equation*}
    \text{Pr}[\mathcal{M}(X) \in \mathcal{S}(Q(X)] \ge \rho,
\end{equation*}
where $\rho$ is the level of confidence that indicating the likelihood of the noisy output falls within $\mathcal{S}(Q(X))$.

\subsection{Privacy Boosting DP Mechanism}\label{sec.mech}

In this section, we present our privacy boosting mechanism. Given the utility constraint in \eqref{eq:utility}, our mechanism consists a kernel DP mechanism $\mathcal{M}$, and reweights its probability density function (PDF) according to \eqref{noise_dist}. For the kernel DP mechanism, one can instantiate any standard noise-adding mechanism that is suitable for the target problem. For example, discrete Gaussian mechanism for discrete domains. The mechanism outputs the noisy query answer as $\widehat{Q(X)}\sim f_{pb}(y|Q(X))$, where $f_{pb}$ is defined as follows.


\begin{equation}\label{noise_dist}
    f_{pb}(y|Q(X)) =
    \begin{dcases} 
        \frac{f_{\mathcal{M}}(y)}{1-\bar{p}_{\mathcal{S}(Q(X))}q}, & \text{if } y\in \mathcal{S}(Q(X)) \\
        \frac{f_{\mathcal{M}}(y)(1-q)}{1-\bar{p}_{\mathcal{S}(Q(X))}q}, & \text{otherwise} 
    \end{dcases}
\end{equation}
where $\bar{p}_{\mathcal{S}(Q(X))}\overset{\Delta}{=}\int_{y\notin{\mathcal{S}(Q(X))}}f_{\mathcal{M}}(y)dy$ is the probability that the output from the kernel mechanism does not fall in the preferred region, and 
\begin{equation}\label{eq:q}
    q=\max_{Q(X)}\frac{1}{\rho}+\frac{1}{\bar{p}_{\mathcal{S}(Q(X))}}-\frac{1}{\rho \bar{p}_{\mathcal{S}(Q(X))}}.
\end{equation}
One can easily verify that this is a valid probability distribution as $\int_{-\infty}^{\infty}f_{pb}(y_n|Q(X))dy_n=1$. We provide the proof in Appendix \ref{app.validdist}.

When $\rho\le \min_{Q(X)}p_{\mathcal{S}(Q(X))}$ ($p_{\mathcal{S}(Q(X))}=1-\bar{p}_{\mathcal{S}(Q(X))}$), which means the kernel mechanism already satisfies the utility constraint and we do not need to do any reweighting to that, our $q$ becomes $0$ and the resulting noise distribution is identical to that in the original kernel mechanism. When $\rho \gets 1$, $q \gets 1$ and the resulting noise distribution is a normalized $f_{\mathcal{M}}$ with output support bounded within $\mathcal{S}(Q(x))$. 

While standard DP mechanisms rely on a fixed noise distribution, the noise distribution of PB-DP mechanisms are carefully designed with: (a) a standard DP noise component that preserves essential properties, such as privacy accounting techniques, and (b) a step function with a boosted region, parameterized by the specified utility boundaries. This design allows PB-DP mechanisms to offer a better tradeoff by introducing new mechanism parameters that can be specifically optimized according to utility constraints. Additionally, sampling a noise instance from the PB-DP noise distribution is efficient. Since we can express the CDP function, which holds true for most standard DP noise-adding mechanisms, inverse transform sampling allows us to efficiently sample the noise. For distributions that are harder to sample, rejection sampling remains a viable option.

\subsection{Privacy Analysis}\label{sec.account}



Fix a pair of neighboring datasets $X,X'$.
The privacy loss random variable of our privacy boosting mechanism is  $\Gamma_{X/X'} \overset{\Delta}{=} \log \frac{\mathcal{M}(X)(y)}{\mathcal{M}(X')(y)}$, where $y\sim \mathcal{M}(X)$. Similarly, we have $\Gamma_{X'/X} \overset{\Delta}{=} \log \frac{\mathcal{M}(X')(y)}{\mathcal{M}(X)(y)}$, where $y\sim \mathcal{M}(X')$. For simplicity, we will only consider $\Gamma_{X/X'}$, and the results will follow directly for $\Gamma_{X'/X}$. We omit the subscript $X,X'$ for the rest of this subsection.
Let $f_\Gamma$ denote the PLD with respect to $\Gamma$. 
We also define the following privacy loss:
\begin{equation}\label{eq:leakage}
\begin{aligned}
\begin{dcases}
&\mathcal{L}_1 \overset{\Delta}{=} \log\left(\frac{{1-\bar{p}_{\mathcal{S}(Q(X'))}q}}{{1-\bar{p}_{\mathcal{S}(Q(X))}q}}\right)\\
&\mathcal{L}_2 \overset{\Delta}{=} -\log(1-q).
\end{dcases}
\end{aligned}
\end{equation}


Let $Z$ denote the privacy loss random variable of kernel DP mechanism, and $f_Z$ be the corresponding PLD.
Then we define a shifted PLD of $f'_Z(z)$:
\begin{equation*}
     f'_Z(z) \overset{\Delta}{=} f_Z(z - \mathcal{L}_1).
\end{equation*}
Denote
\begin{equation}\label{eq:tau}
    \tau_u \overset{\Delta}{=} \sup \mathcal{S}(Q(X)), ~\tau_l \overset{\Delta}{=} \inf \mathcal{S}(Q(X)).
\end{equation}
Then the PLD of the PB-DP mechanism corresponds to the following theorem.

\begin{thm}\label{thm:PLD_BR-DP}
The PLD of our privacy boosting mechanism for a pair of neighboring datasets $X, X'$, given the PLD of the kernel DP mechanism $f_Z$ can be represented as:
\begin{equation*}
\begin{aligned}
    &f_{\Gamma}(\gamma) = 
 W_1 f'_Z(\gamma - \mathcal{L}_2)+ W_2 f'_Z(\gamma + \mathcal{L}_2) +W_3f'_Z(\gamma).
\end{aligned}
\end{equation*}
where 

\begin{equation}\label{eq:W}
    \begin{aligned}
    \begin{dcases} 
        &W_1 = \int_{\min\{ \tau_u, \tau'_u \}}^{\max\{ \tau_u, \tau'_u \}} f_{\mathcal{M}(X)}(y) dy;\\
        &W_2 = \int_{\min\{ \tau_l, \tau'_l \}}^{\max\{ \tau_l, \tau'_l \}} f_{\mathcal{M}(X)}(y) dy; \\
        &W_3 = 1 -W_1 -W_2.
    \end{dcases}
\end{aligned}
\end{equation}

\end{thm}


Our mechanism first introduces an additional privacy loss of $\mathcal{L}_1$ to the kernel DP mechanism's existing privacy loss, causing a shift in the PLD. This shift results from the different likelihoods of $\bar{p}_{\mathcal{S}(Q(X'))}$ and $\bar{p}_{\mathcal{S}(Q(X))}$ due to the potential discrepancy of $\mathcal{S}(Q(X'))$ and $\mathcal{S}(Q(X))$. Depending on the region where an output $y$ falls in, the mechanism incurs one of two types of privacy leakages: $\mathcal{L}_2$ and $-\mathcal{L}_2$, 
corresponding to two events $\{y \in \mathcal{S}(Q(X)); y \notin \mathcal{S}(Q(X'))\}$, and $\{y \in \mathcal{S}(Q(X')); y \notin \mathcal{S}(Q(X))\}$, respectively. The probabilities of the two events are denoted by $W_1$ and $W_2$, respectively,  when $y \sim \mathcal{M}(X)$.





We can use the PLD to characterize the standard $(\epsilon,\delta)$-DP. 
The following proposition shows the privacy profile of a PB-DP mechanism. Let $\delta_Z(\epsilon)$ denote privacy profile when the privacy loss random variable is $Z$.

\begin{prop}\label{thm.profile}
Given the shifted privacy profile of the kernel DP mechanism $\delta'_Z(\epsilon)\overset{\Delta}{=} \delta_Z(\epsilon-\mathcal{L}_1)$, the privacy profile of the PB-DP mechanism is as follows:
\begin{equation*}
\begin{aligned}
    &\delta_{\Gamma}(\epsilon) = \max_{X,X'}\{
   W_1 \delta'_Z(\epsilon - \mathcal{L}_2)+ W_2\delta'_Z(\epsilon + \mathcal{L}_2)+ W_3 \delta'_Z(\epsilon) \}.
\end{aligned}
\end{equation*}
\end{prop}
Proposition \ref{thm.profile} suggests that the privacy profile of our mechanism is a linear combination of the privacy profile of kernel DP evaluated at different privacy leakages weighted at their probabilities of occurring.

Given the property of the PLD, we can also measure the privacy leakage of our mechanism captured by RDP.
\begin{prop}\label{thm.rdp}
Given a kernel DP mechanism that satisfies $(\alpha,\epsilon_0)$-RDP, the PB-DP mechanism is $(\alpha,\epsilon)$-RDP for 
\begin{equation*}
\begin{aligned}
    \epsilon =& \epsilon_0 + \max_{X,X'}\left\{\mathcal{L}_1+\frac{1}{\alpha - 1}\log\left[ W_1e^{(\alpha -1)\mathcal{L}_2} \right.\right. \\
    &~~~~~~~~~~~~~~~~~~~~~~~~~~~~~~~\left.\left.+ W_2e^{-(\alpha-1)\mathcal{L}_2} + W_3 \right]\right\}.
\end{aligned}
\end{equation*}
\end{prop}

We define the dominating pair $X,X'$ is the pair that maximizes the expression above. We note that this pair is not necessarily the worst-case neighboring datasets for the entire PB-DP mechanism, but rather the pair that maximizes the privacy leakage caused by the boosting part. This pair may differ from the worst-case pair for the kernel DP mechanism. In Proposition \ref{thm.profile} and Theorem \ref{thm.rdp}, we upper bound the privacy leakage caused by the kernel DP mechanism using its privacy-loss parameters. 
Additionally, the dominating pair $X,X
$ depends solely on the preferred region and the boosting parameter $q$, and is independent of the choice of the kernel DP mechanism. In subsequent sections, determining the privacy loss relies on finding this dominating pair, and we will demonstrate that such a pair can always be found. It is also worth noting that in the literature, the term ``dominating pair" typically refers to a pair of distributions, not necessarily a pair of datasets. We slightly abuse the notation here for clarity. 

We also note that our PB-DP mechanisms may not be the ideal choice for pure DP scenarios. This is because the soft-bounded design in the PB-DP noise distribution introduces non-negative additional privacy loss ($\mathcal{L}_1, \mathcal{L}_2$)  with certain probabilities ($W_1, W_2$). With certain relaxations, these leakages are accounted for as ordered expectations, weighted by probabilities that are typically very small. However, in the context of pure DP accounting, the maximum leakage is increased by these leakages, regardless of how small the probabilities are.

\subsection{Privacy Accounting for Sequential Composition}\label{sec.comp}

Observe that the  PLD in Theorem \ref{thm:PLD_BR-DP} can be rewritten as a convolution of two privacy loss distributions:
\begin{equation*}
    f_{\Gamma}(\gamma)=f'_Z(\gamma)\ast f_R (\gamma).
\end{equation*}
Here $\ast$ denotes the convolution operation, and $f'_Z$ is the shifted PLD of the kernel DP mechanism,  and $f_R$ is a privacy loss distribution defined as follows.
\begin{equation*}
\begin{aligned}
    f_{R}(r) =& W_1 f_{\text{Dirac}}(r - \mathcal{L}_2)+W_2 f_{\text{Dirac}}(r+\mathcal{L}_2)+ W_3 f_{\text{Dirac}}(r),
\end{aligned}
\end{equation*}
where $f_{\text{Dirac}}$ is the Dirac function defined such that $f_{\text{Dirac}}(t) = 1$, iff $t =0$, otherwise   $f_{\text{Dirac}}(t) = 0$. The Dirac function represents a distribution where the entire probability mass is concentrated at 0. This means that with probability 1, the privacy loss is 0, making it a valid distribution for modeling privacy loss. The function $f_R$ thus represents a privacy loss distribution where the privacy loss can take on values of $\mathcal{L}_2$, $-\mathcal{L}_2$, or 0, with respective probabilities $W_1, W_2,$ and $W_3$.

Then, the following theorem describes the PLD of a PB-DP mechanism after $T$-fold homogeneous compositions:


\begin{thm}
    The privacy loss distribution after $T$-fold homogeneous composition of the PB-DP mechanism with PLD of $f_{\Gamma}(\gamma)$ is: 
    \begin{equation}
    \begin{aligned}
    f^T_{\Gamma}(\gamma)=&\sum_{e_1+e_2\le T}\binom{T}{e_1,e_2} W_1^{e_1}W_2^{e_2}(1-W_1-W_2)^{T-e_1-e2}\\
    &~~~~~~~\cdot f'_Z \ast^T f'_Z (\epsilon - (e_1-e_2)\mathcal{L}_2).
        \end{aligned}
    \end{equation}
     where the coefficients $e_1,e_2$ are non-negative integers.
\end{thm}


    
Then, we have the following proposition for the privacy profile.
\begin{prop}\label{thm:compo}
    The T-fold homogeneous composition of a PB-DP mechanism is $(\epsilon,\delta)$-DP for 
    \begin{equation*}
    \begin{aligned}
    \delta(\epsilon)=&\sum_{e_1+e_2\le T}\binom{T}{e_1,e_2} W_1^{e_1}W_2^{e_2}(1-W_1-W_2)^{T-e_1-e_2}\\
    &~~~~~~~\cdot \delta_Z^{\prime T} (\epsilon - (e_1-e_2)\mathcal{L}_2).
    \end{aligned}
    \end{equation*}
 where $\delta_Z^{\prime T}(z)$ denotes the shifted privacy profile of the kernel DP mechanism after $T$-fold homogeneous composition.
\end{prop}


We next present an accounting algorithm to capture the privacy loss parameter for $T$- fold homogeneous composition of PB-DP in Algorithm 1, which efficient releases $\delta$ for a given $\epsilon$. As $\mathcal{L}_2$ and $-\mathcal{L}_2$ are symmetric about $0$, after a $T$-fold convolution, there are $2T + 1$ possible leakages, each with an increment of $\mathcal{L}_2$. The probability of each leakage can be recursively calculated by searching over all possible combinations of $W_1$s and $W_2$s that achieving the corresponding leakage. 

 \begin{algorithm}
\caption{Composition accountant for PBDP }
\hspace*{\algorithmicindent}
\textbf{Input:} $q$, $\epsilon_0$, $\delta_y$, target $\epsilon$, $T$\\
 \hspace*{\algorithmicindent} \textbf{Output:} $\delta(\epsilon)$.
\begin{algorithmic}[1]
\State Determine dominating pair of $X$ and $X'$;
\State Get $\{\tau_l, \tau_u, \tau'_l, \tau'_u\}$ with $\mathcal{S}(Q(X))$ and $\mathcal{S}(Q(X'))$;
\State Get $q$ $\gets$ $\eqref{eq:q}$;
\State Determine $\mathcal{L}_1, \mathcal{L}_2$ $\gets$ \eqref{eq:leakage};
\State Determine $W_1$, $W_2$ $\gets$ \eqref{eq:W};
\State Initialize vector $\mathcal{V}$ with length of $2T+1$;
\State Initialize $\delta = 0$;
\For{$e_1$ in range$(T+1)$}
\For{$e_2$ in range$(T - e_1 + 1)$}
\State $u=\binom{T}{e_1}\binom{T-e_1}{e_2}W_1^{e_1}W_2^{e_2}(1-W_1-W_2)^{T-e_1-e_2}$;
\State $\mathcal{V}[e_1 -e_2 - 1 +T] \gets \mathcal{V}[e_1 -e_2 - 1 +T] + u$;
\EndFor
\EndFor
\For{$i$ in range (0, 2T + 2)}
\State $\delta \gets \delta + \delta^{\prime T}_Z(\epsilon - \mathcal{V}[i - T])$;
\EndFor\\
\Return $\delta$;
\end{algorithmic}
\label{algo1}
\end{algorithm}

\begin{rmk}
The computational complexity of Algorithm 1 is $\mathcal{O}(T^2)$. 
\end{rmk}

The composition analysis for RDP based PB-DP mechanism is also straightforward, which is given in the following remark.

\begin{rmk}
    For a sequence of PB-DP mechanisms, each satisfying $(\alpha,\epsilon_i)$-RDP, the composition of these mechanisms is $(\alpha, \sum_{i = 1} ^T \epsilon_i)$-RDP.
\end{rmk}

\subsection{Optimal Parameters}\label{sec.optparam}


We next provide a efficient method to search for the smallest privacy-loss parameter such that our PB-DP mechanism satisfies the utility constraint in \eqref{eq:utility}.

For simplicity, we fix $\delta$ or $\alpha$ for the kernel DP mechanism and the entire PB-DP mechanism when measuring privacy. However, it is straightforward to adjust $\delta_0$ and $\alpha$ according to specific application requirements.



To enhance the utility-privacy tradeoff, a larger $\epsilon_0$ in the kernel DP mechanism increases the likelihood of falling in the preferred region $p_{\mathcal{S}(Q(X))}$, and therefore requires a smaller boosting parameter $q$. To summarize, while the privacy loss associated with the kernel DP mechanism increases, the privacy loss incurred by the discrepancy in the boosting region would be smaller. Thus, the optimal PB-DP mechanism embodies a tailored privacy budget allocation between the kernel DP and the boosting part.

We can numerically search for the optimal $\epsilon_0$ that minimizes the total privacy-loss parameter $\epsilon$. As mentioned above, the optimal $\epsilon_0$ yields the optimal privacy budget allocation between the kernel DP and the boosting part. Consequently, there will be a single peak of $\epsilon_0$ that minimizes the total privacy loss. Ternary search is an efficient algorithm for finding the peak of a convex or concave function, which well suits our setting. The steps to find the optimal $\epsilon_0$ are detailed in algorithm 2.

 \begin{algorithm}
\caption{Find Optimal $\epsilon_0$ using Ternary Search}
\hspace*{\algorithmicindent}
\textbf{Input:} $\rho$, $\delta$, ($\alpha$), $\Delta_f$, $X$, $X'$, tol.\\
 \hspace*{\algorithmicindent} \textbf{Output:} Optimal  $\epsilon_0$.
\begin{algorithmic}[1]
    \State $\epsilon_{low} \gets 0$, $\epsilon_{up} \gets \epsilon_{\max}$;
    \State $\tau_l$, $\tau_u$, $\tau'_l$, $\tau'_u$ $\gets$ Eq. \eqref{eq:tau} ($X$, $X'$);
    \While{$\epsilon_{up} - \epsilon_{low} > \text{tol}$}
        \State $\epsilon_1 \gets \epsilon_{low} + \frac{\epsilon_{up} - \epsilon_{low}}{3}$, $\epsilon_2 \gets \epsilon_{up} - \frac{\epsilon_{up} - \epsilon_{low}}{3}$ ;
        \State Get $q_1, q_2$ via \eqref{eq:q} corresponding to ($\epsilon_1,\epsilon_2$);
        \State $\mathcal{L}^1_1, \mathcal{L}^1_2$ $\gets$ Eq. \eqref{eq:leakage} ($\epsilon_0 = \epsilon_1, q_1, X, X'$);
        \State $\mathcal{L}^2_1, \mathcal{L}^2_2$ $\gets$ Eq. \eqref{eq:leakage} ($\epsilon_0 = \epsilon_2, q_2, X, X'$);
        \State $W^1_1, W^1_2$ $\gets$ Eq. \eqref{eq:W} ($\epsilon_0 = \epsilon_1$, $\mathcal{S}(Q(X))$, $\mathcal{S}(Q(X'))$)
        \State $W^2_1, W^2_2$ $\gets$ Eq. \eqref{eq:W} ($\epsilon_0 = \epsilon_2$, $\mathcal{S}(Q(X))$, $\mathcal{S}(Q(X'))$)
        \State $(\epsilon'_1,\delta) \gets$ proposition 1  ($\mathcal{L}^1_1, \mathcal{L}^1_2, W^1_1, W^1_2$);
        \State $(\epsilon'_2,\delta) \gets$ proposition 1  ($\mathcal{L}^2_1, \mathcal{L}^2_2, W^2_1, W^2_2$);
        \State or
        \State $(\alpha, \epsilon'_1) \gets$ proposition 2  ($\mathcal{L}^1_1, \mathcal{L}^1_2$, $W^1_1, W^1_2$);
        \State $(\alpha, \epsilon'_2) \gets$ proposition 2  ($\mathcal{L}^2_1, \mathcal{L}^2_2$, $W^2_1, W^2_2$);
        \If{$ \epsilon'_1>\epsilon'_2$}
            \State $\epsilon_{low} \gets \epsilon_1$;
        \Else
            \State $\epsilon_{up} \gets \epsilon_2$;
        \EndIf
    \EndWhile
    \State \Return $(\epsilon_{up} + \epsilon_{low})/ 2$
\end{algorithmic}
\end{algorithm}

Given the total privacy budget $\epsilon_{max}$, to find the optimal $\epsilon_0$, we start by initializing $\epsilon_{low}$ to 0 and $\epsilon_{up}$ to $\epsilon_{max}$. During the iterative search, while the difference between $\epsilon_{low}$ and $\epsilon_{up}$ is greater than the tolerance, we divide the interval by setting $\epsilon_1 \gets \epsilon_{low} + \frac{\epsilon_{up} - \epsilon_{low}}{3}$, $\epsilon_2 \gets \epsilon_{up} - \frac{\epsilon_{up} - \epsilon_{low}}{3}$. For $\epsilon_{low}$ and $\epsilon_{up}$, we then calculate the corresponding $q_1$ and $q_2$, and determine the total privacy losses $\epsilon'_1$ and $\epsilon'_2$ using Theorem \ref{thm.profile} for $(\epsilon,\delta)$-DP or Theorem \ref{thm.rdp} for RDP. Depending on the results, if $\epsilon'_1>\epsilon'_2$, we update $\epsilon_{low}=\epsilon_1$; otherwise, we set $\epsilon_{up}=\epsilon_2$. The optimal $\epsilon_0$ is obtained as $(\epsilon_{low}+\epsilon_{up})/2$.

We note that the search algorithm for determining the optimal PB-DP parameters based on utility constraints can be executed offline. Once the output domain is defined, these parameters—and consequently the noise distribution—are fixed. It's important to note that optimal searching is optional; even without it, one can still achieve improved utility-privacy trade-offs by simply sampling a noise instance from the PB-DP noise distribution.

\begin{figure*}[t]
\centering 
\subfigure[Data dependent preferred region (relative error)]
{\includegraphics[width=0.32\textwidth]{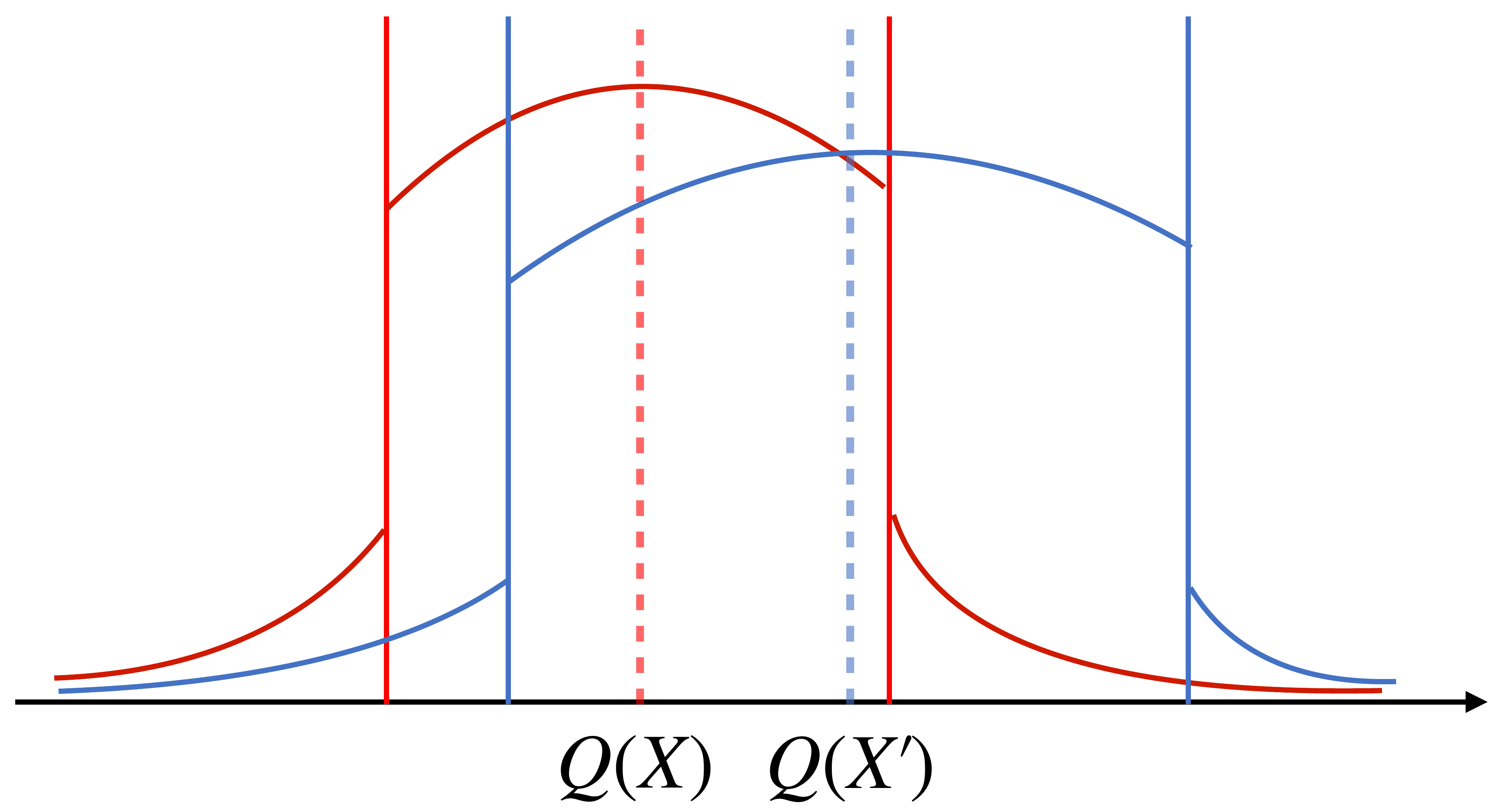}
\label{dep_noise}}
\subfigure[Fixed preferred output region]
{\includegraphics[width=0.32\textwidth]{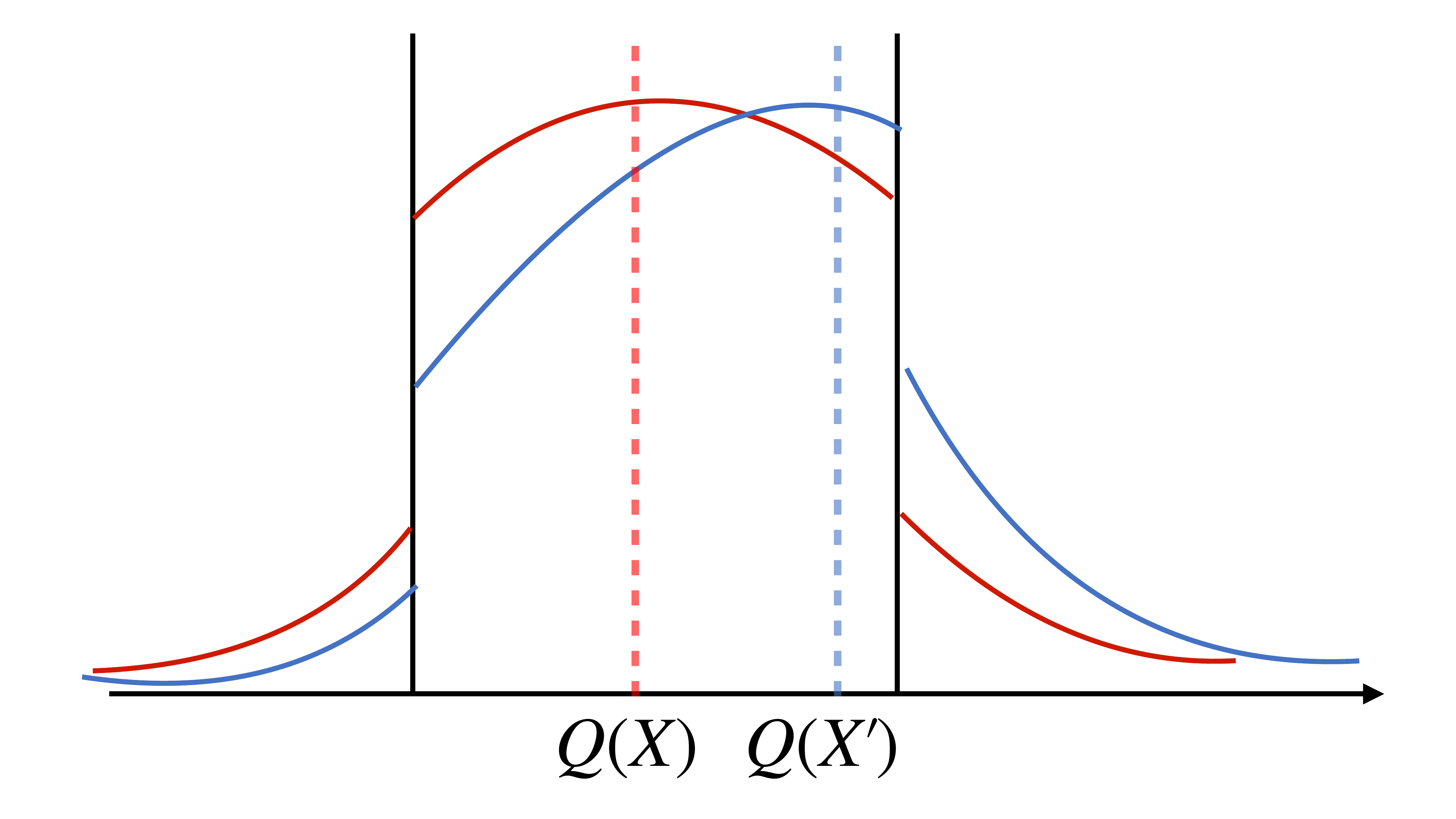}
\label{fix_bound}}
\subfigure[Data independent preferred region (absolute error)]
{\includegraphics[width=0.32\textwidth]{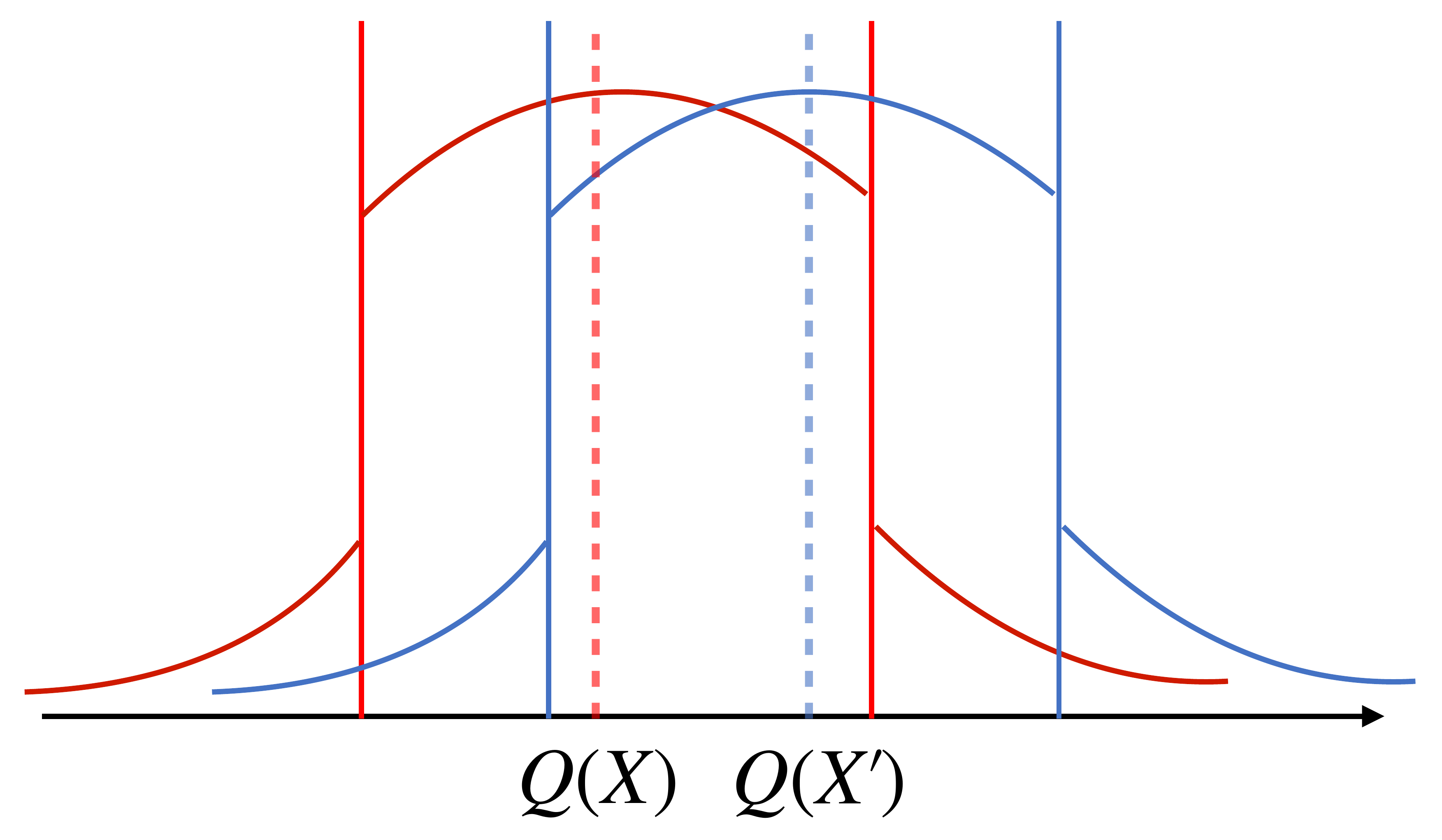}
\label{ind_noise}}
\caption{Illustration of privacy boosting mechanisms for the three special cases studied in this paper. We take one dimensional Gaussian kernel as an example. In each figure, $Q(X)$ and $Q(X')$ denotes true aggregations from neighboring datasets. Solid vertical lines represent boundaries of preferred regions for each case.}
\end{figure*}

\section{Case Study: Mechanisms with Different Types of preferred regions}

In this section, we present case studies demonstrating how the privacy boosting mechanism can be adapted for three different specifications of $\{\mathcal{S}(Q(X))\}$  First, we illustrate a data-dependent preferred region using constrained relative error as an example. Next, we consider a case with fixed preferred region. Then, we examine a case with a data-independent preferred region using absolute error as a constraint. Finally, we explore a boosted randomized response in the local model.

\subsection{Mechanisms with data dependent preferred regions}

A direct application of our mechanism is that the preferred noisy region depends on the true query answer, i.e., $\mathcal{S}(Q(X))$ varies for different $Q(X)$. Specifically, we consider a relative error bound where the preferred noise region is defined as: $$\mathcal{S}(Q(X)) \overset{\Delta}{=} \left\{y: ||y-Q(X)||_l\le \theta ||Q(X)||_l + \tau\right\}.$$ Here, $\theta\in[0,1]$ defines the relative error ratio, and $\tau > 0$ is an offset that ensures the preferred region remains valid.  

We use a one-dimensional query answer as an example, though our analysis can be extended to multi-dimensional answers.  For relative error bounds, let  $\Phi_{\mathcal{M}}$ denote the CDF of noise in the kernel DP mechanism centered at $0$. The probability $p_{\mathcal{S}(Q(X))}$ can be specified as
\begin{equation*}
\Phi_{\mathcal{M}}(\theta |Q(X)| + \tau)-\Phi_{\mathcal{M}}(-\theta |Q(X)| - \tau).
\end{equation*}
The minimal $p_{\mathcal{S}(Q(X))}$ occurs when $Q(X) = 0$, yielding:
\begin{equation*}
    \min p_{\mathcal{S}(Q(X))} = p_{\mathcal{S}(0)} = \Phi_{\mathcal{M}}(\tau)-\Phi_{\mathcal{M}}( - \tau).
\end{equation*}
The corresponding $q$ can be derived as:
\begin{equation*}
    q=\frac{1}{\rho}+\frac{1}{1-p_{\mathcal{S}(0)}}-\frac{1}{\rho (1-p_{\mathcal{S}(0)})}.
\end{equation*}
The parameters in the PLD expression can then be specified according to the following proposition.

\begin{prop}

For a preferred region defined by relative error, the privacy losses defined in \eqref{eq:leakage} can be specified as follows:
\begin{small}
\begin{equation*}
    \begin{aligned}
    \begin{dcases}
        &\mathcal{L}_1 = \log\left(\frac{1 - (1-\Phi_{\mathcal{M}}(\theta\Delta_Q+\tau) +\Phi_{\mathcal{M}}(-\theta\Delta_Q - \tau))q}{1-(1-\Phi_{\mathcal{M}}(\tau_u) + \Phi_{\mathcal{M}}(\tau_l))q}\right),\\
        &\mathcal{L}_2 = -\log\left(1-q\right),\\
    \end{dcases}
    \end{aligned}
    \end{equation*}
\end{small} 
where $\Delta_Q$ is the sensitivity of the query.
The corresponding probabilities defined in \eqref{eq:W} are:
 \begin{equation*}
\begin{aligned}
\begin{dcases}
  &W_1 =\Phi_{\mathcal{M}}(\theta\Delta_Q + \tau) - \Phi_{\mathcal{M}}(\tau),\\
  &W_2 = \Phi_{\mathcal{M}}(\theta\Delta_Q - \tau) - \Phi_{\mathcal{M}}(-\tau).
\end{dcases}
\end{aligned}
\end{equation*}

\end{prop}
Intuitively, the largest  $\mathcal{L}_1$ is achieved at greatest discrepancy between $p_{\mathcal{S}(Q(X))}$ and $p_{\mathcal{S}(Q(X'))}$, which corresponds to the case where $Q(X) = 0$, $Q(X') = \Delta_Q$. 


For data-dependent utility bounds, all privacy losses defined in \eqref{eq:leakage} and all the probabilities defined in \eqref{eq:W} are non-zero. This means the determination of optimal parameters in Algorithm 2 and the privacy accounting algorithms cannot be further simplified. In the next section, we introduce other special cases where some of the privacy losses or probabilities are zero, simplifying the process.

\subsection{Mechanism with fixed preferred output region}

Consider a fixed preferred output region $\mathcal{S}$ that does not dependent on the true answer $Q(X)$. For example, a valid answer must be within a certain range. 
Specifically, we define $\mathcal{S}(Q(X))\overset{\Delta}{=}\{y:  \tau_l \le y \le \tau_u \}$.

For one dimensional data, the probability of falling within the preferred region is given by:
$$p_{\mathcal{S}(Q(X))} = \Phi_{\mathcal{M}}(\tau_u - Q(X)) - \Phi_{\mathcal{M}}(\tau_l- Q(X)).$$
The minimal $p_{\mathcal{S}(Q(X))}$ is reached at the edge of the region. 
For simplicity, we assume the noise distribution in the kernel mechanism is symmetric.
We consider $Q(X) = \tau_l$ and $Q(X')=\tau_l+\Delta_Q$.  Thus,
\begin{equation*}
    \min p_{\mathcal{S}(Q(X))} = p_{\mathcal{S}(\tau_l)} = \Phi_{\mathcal{M}}(\tau_u - \tau_l) - \Phi_{\mathcal{M}}(0).
\end{equation*}
The boosting rate is then determined as:

\begin{equation*}
    q=\frac{1}{\rho}+\frac{1}{1-p_{\mathcal{S}(\tau_l)}}-\frac{1}{\rho (1-p_{\mathcal{S}(\tau_l)})}.
\end{equation*}
The parameters in the PLD expression are specified in the following proposition.

\begin{prop}
    For a fixed preferred output region, the privacy losses defined in \eqref{eq:leakage} are:
    \begin{equation*}
    \begin{aligned}
        &\mathcal{L}_1 = \log\left(\frac{1-(1-\Phi_{\mathcal{M}}(\tau_u-\tau_l-\Delta_Q) + \Phi_{\mathcal{M}}(-\Delta_Q))q}{1-(1-\Phi_{\mathcal{M}}(\tau_u-\tau_l) + \Phi_{\mathcal{M}}(0))q}\right);
    \end{aligned}
    \end{equation*}
$\mathcal{L}_2$ does not exist in this case as the corresponding probabilities defined in \eqref{eq:W} are:
\begin{equation*}\label{eq:W_fix}
    \begin{aligned}
    \begin{dcases} 
        &W_1 = 0,\\
        &W_2 = 0.\\
    \end{dcases}
\end{aligned}
\end{equation*}
\end{prop}

For fixed preferred output region, $\mathcal{L}_1$ achieves its maximum when $X$ and $X'$ are one at the boundary and one shifted by $\Delta_f$. Since the preferred region is fixed for all possible answers, the support misalignment probability is zero, resulting in zero values for both $W_1$ and $W_2$. Therefore, $\mathcal{L}_2$ does not exist in this case. In essence, our mechanism adds an additional privacy loss of $\mathcal{L}_1$ to the kernel DP mechanism. Compared to bounded mechanisms, such as \cite{Geng2018TruncatedLM, 2022-2-7-0193, holohan2018bounded}, which forces $q = 1$, our mechanism provides more flexibility by varying $q$ to adjust $\mathcal{L}_1$. This flexibility enlarges the feasible region of the $\epsilon$, $\delta$ tradeoff by reducing the achievable $\epsilon$ for any given $\delta$. 

With $W_1 = W_2 = 0$, the privacy analysis becomes more straightforward.
\begin{rmk}
The PB-DP mechanism with fixed output region achieves $(\epsilon,\delta)$-DP, where $$\delta = \delta_Z(\epsilon-\mathcal{L}_1),$$ and $\delta_Z(\epsilon)$ is the
privacy profile of the kernel DP mechanism. On the other hand, it also achieves $(\alpha, \epsilon_0+\mathcal{L}_1)$-RDP, when the kernel DP mechanism is $(\alpha,\epsilon_0)$-RDP.
\end{rmk}
For $T$-fold homogeneous composition, the PLD becomes $$f^{T}_{\Gamma}(\gamma) = f^T_{Z} (\gamma - T\mathcal{L}_1),$$ where $f_Z^T$ denotes the $T$-fold homogeneous composed PLD of the kernel DP mechanism.


In section \ref{sec.expfixed}, we numerically compare the feasible regions of privacy parameters $\epsilon, \delta$ of our PB-DP mechanism and traditional bounded DP mechanisms.

\subsection{Mechanism with data-independent preferred region}

Next, we consider a scenario where the preferred region is data-independent and depends only on the noise magnitude. Specifically, we define the preferred region as 
$\mathcal{S} \overset{\Delta}{=}\{y: ||y-Q(X)||_l \le \tau\}$, where $\tau\in[0,\infty)$. 
In this case, 
\begin{equation}
    p_{\mathcal{S}(Q(X))}=p_{\mathcal{S}} = \Phi_{\mathcal{M}}(\tau) - \Phi_{\mathcal{M}}(-\tau),
\end{equation}
which is independent of the true query answer $Q(X)$.
The corresponding $q$ is 
\begin{equation*}
    q = \frac{\rho - p_{\mathcal{S}}}{\rho (1-p_{\mathcal{S}})}.
\end{equation*}
The additional privacy losses can be specified as follows. 
 
\begin{prop}\label{PLD_BR-DP}
For preferred region defined by the absolute error, the privacy losses defined in \eqref{eq:leakage} are:
\begin{equation*}
\begin{aligned}
\begin{dcases}
    &\mathcal{L}_1 = 0;\\
    &\mathcal{L}_2 = -\log\left({1-q}\right).\\
\end{dcases}
\end{aligned}
\end{equation*}
The corresponding probabilities defined in \eqref{eq:W} are
\begin{equation}\label{eq:W_fix}
    \begin{aligned}
        &W_1 = W_2 = \Phi_{\mathcal{M}}(-\tau +\Delta_Q) - \Phi_{\mathcal{M}}(-\tau);\\
\end{aligned}
\end{equation}
\end{prop}
For a data-independent preferred region, the noisy query answer $\widehat{Q(X)}$ have the same probability of being released within the preferred region, regardless of the query answer $Q(X)$. While $\bar{p}_{\mathcal{S}(Q(X))} = \bar{p}_{\mathcal{S}(Q(X'))}$, and are boosted with identical rate $q$. This results in a $\mathcal{L}_1=0$. Additionally, the additional privacy loss $\mathcal{L}_2$ caused by preferred region misalignment is a constant and solely determined by the boosting rate $q$. The probabilities of incurring $\mathcal{L}_2, -\mathcal{L}_2$ are identical. These parameters are all data-independent, and thus does not require us to find a dominant pair $X$, $X'$. We then have the following statement for the privacy guarantees.

\begin{rmk}
     The privacy boosting mechanism with kernel DP mechanism that has privacy profile $\delta_Z$ under absolute error constraint is $(\epsilon,\delta)$-DP for 
     \begin{equation*}
         \delta = W_1 [\delta_Z(\epsilon - \mathcal{L}_2) +  \delta_Z(\epsilon + \mathcal{L}_2)] + (1-2W_1) \delta_Z(\epsilon).
     \end{equation*}
    When instantiating a $(\alpha,\epsilon_0)$-RDP kernel mechanism, it is $(\alpha,\epsilon)$-RDP for 
    \begin{equation*}
        \epsilon = \epsilon_0 +\frac{1}{\alpha -1}\log \left\{1-2W_2 + W_2 e^{(\alpha-1)}(e^{\mathcal{L}_2} + e^{-\mathcal{L}_2})\right\}.
    \end{equation*}
\end{rmk}


\subsection{PB-Local DP with General Randomize Response}\label{sec:pb-grr}

In this section, we consider discrete data types. Unlike previous cases, we consider a local model with pure $\epsilon$-Local DP (LDP) guarantee ($\delta = 0$). We explore a scenario where there are preferred output regions for each data point, which can be deterministic, such as categories, or rotating, such as neighboring numbers for ranking. Specifically, consider a finite data support $\mathcal{X}$ with cardinally $|\mathcal{X}|$. Each data point $X\in \mathcal{X}$ has a preferred region $\mathcal{S}(X)$. For example, for data with certain class labels, perturbing an item's label to another within the same category is more accurate than perturbing it to a different category. Another example is ranking data: perturbing a rating score from $1$ to $2$ is preferable to perturbing it to $9$.

\subsubsection{Privacy Boosting General Randomize Response}

We use the general randomize response mechanism (GRR) as the kernel mechanism. 
Let $p$ denote the probability that the data is truthfully reported; $p_s$ denote the probability that the data is perturbed to another item in the same class or in its neighbor; $p_{\bar{s}}$ denote the probability that the data is perturbed to another item in another class or outside its neighbor. Then the perturbation parameters of our privacy boosting mechanism follows the following proposition.

\begin{prop}
    The privacy boosting mechanism that achieves $\epsilon$-LDP ($\epsilon$-PB-LDP) with bounded preferred region with size $|\mathcal{S}|$ for all $x\in\mathcal{X}$ can be specified as
    \begin{equation*}
    \begin{aligned}
    \begin{dcases}
        &p = e^{\epsilon}/(e^{\epsilon} + (|\mathcal{S}| -1) e^{\epsilon-\epsilon_0} + |\mathcal{X}| - |\mathcal{S}|);\\
        &p_s = e^{\epsilon-\epsilon_0}/(e^{\epsilon} + (|\mathcal{S}| -1) e^{\epsilon-\epsilon_0} + |\mathcal{X}| - |\mathcal{S}|);\\
        &p_{\bar{s}} =  1/(e^{\epsilon} + (|\mathcal{S}| -1) e^{\epsilon-\epsilon_0} + |\mathcal{X}| - |\mathcal{S}|).\\
    \end{dcases}
    \end{aligned}
    \end{equation*}
\end{prop}

\begin{rmk}
The confidence $\rho$ corresponds to the PB-LDP is:
\begin{equation*}
    \rho = \frac{e^{\epsilon} + (|\mathcal{S}| - 1)e^{\epsilon-\epsilon_0}}{e^{\epsilon} + (|\mathcal{S}| -1) e^{\epsilon-\epsilon_0} + |\mathcal{X}| - |\mathcal{S}|}.
\end{equation*}
\end{rmk}

\begin{figure*}[t]
\centering 
\subfigure[Approximate DP {$\rho = 0.9, \Delta_Q = 1$.}]
{\includegraphics[width=0.49\textwidth]{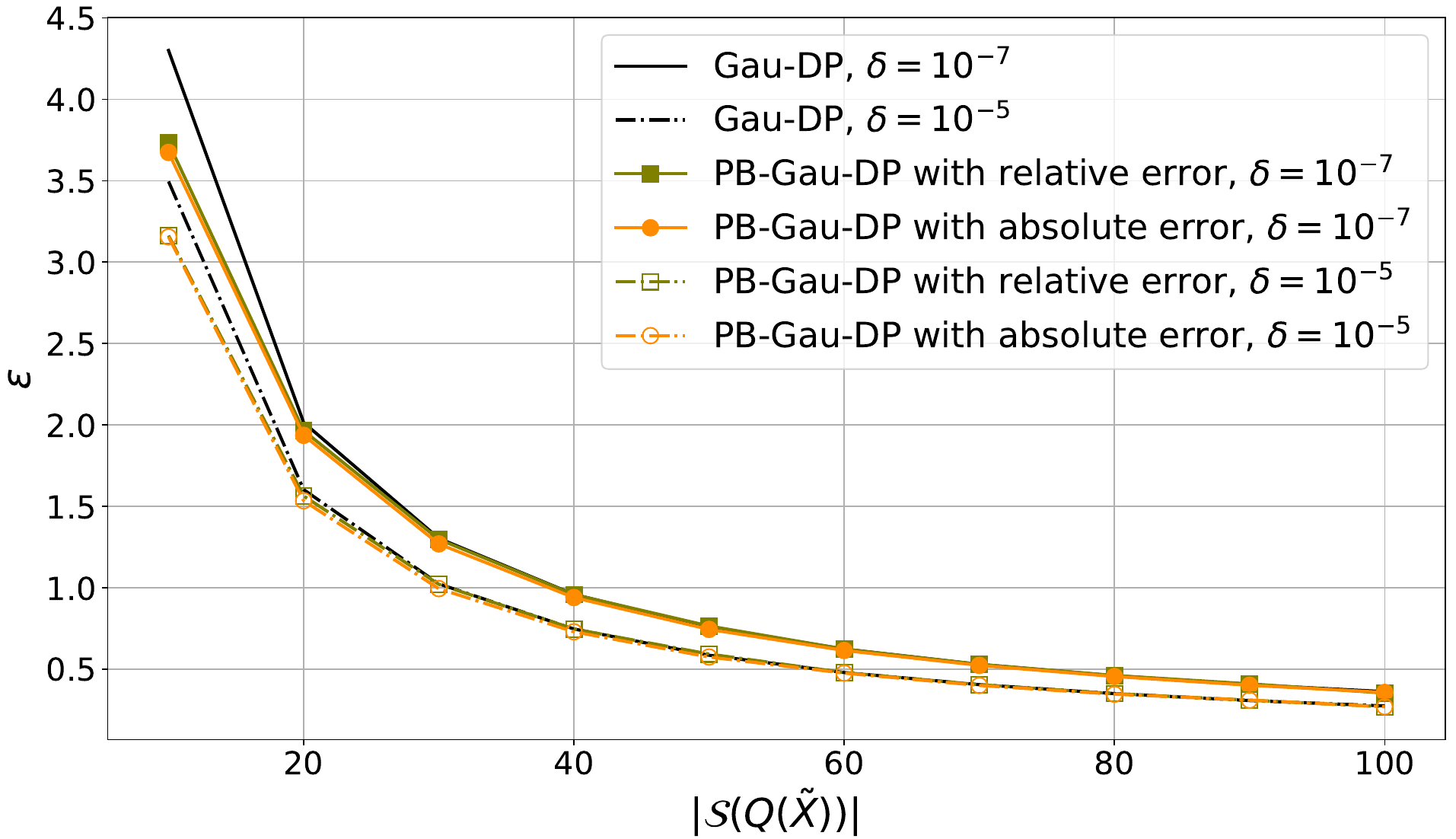}
\label{fig:DP_privacy1}}
\subfigure[Approximate DP {$\rho = 0.8, \Delta_Q = 4$.}]
{\includegraphics[width=0.49\textwidth]{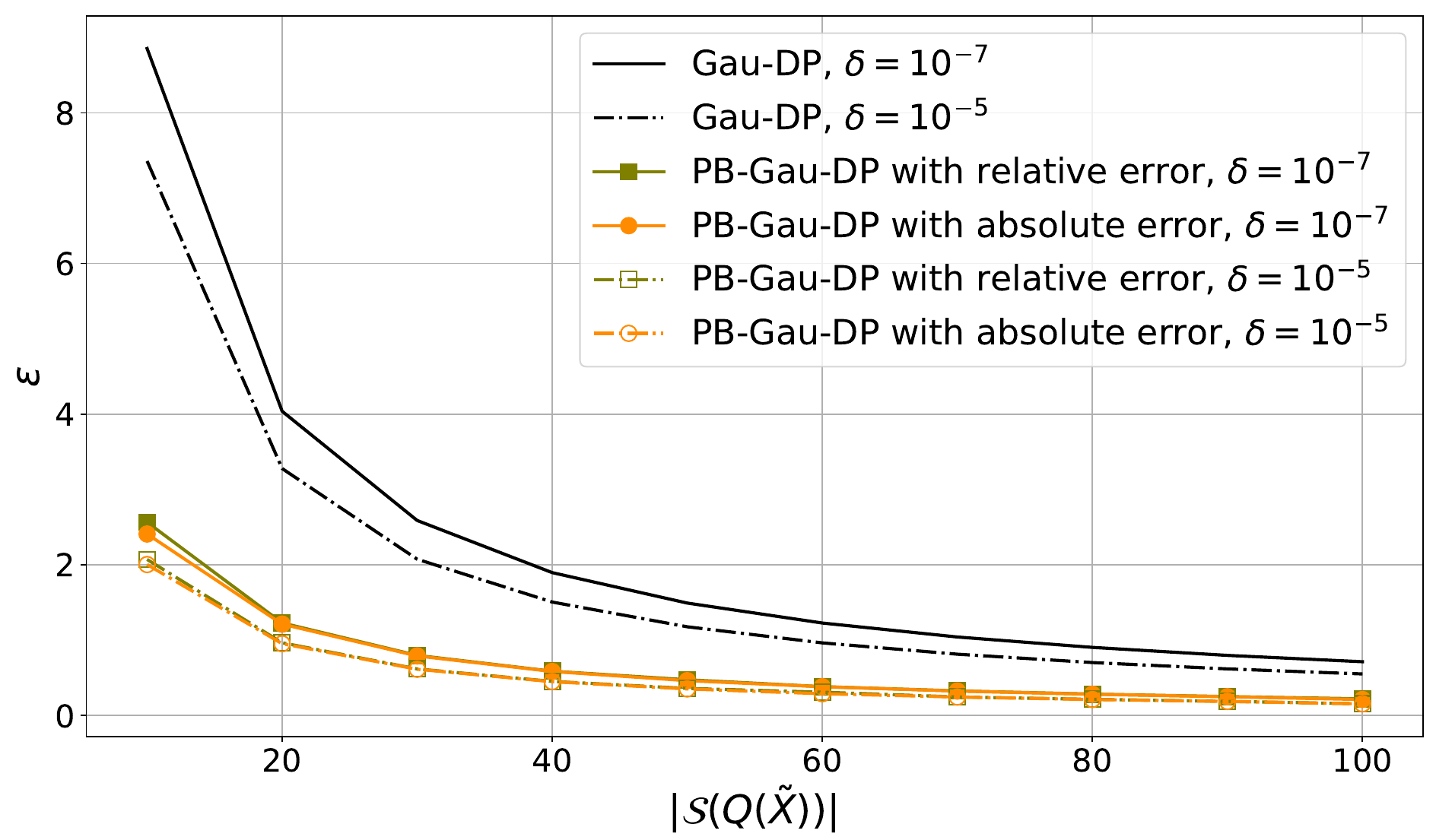}
\label{fig:DP_privacy2}}
\subfigure[RDP {$\rho = 0.9, \Delta_Q = 1$.}]
{\includegraphics[width=0.49\textwidth]{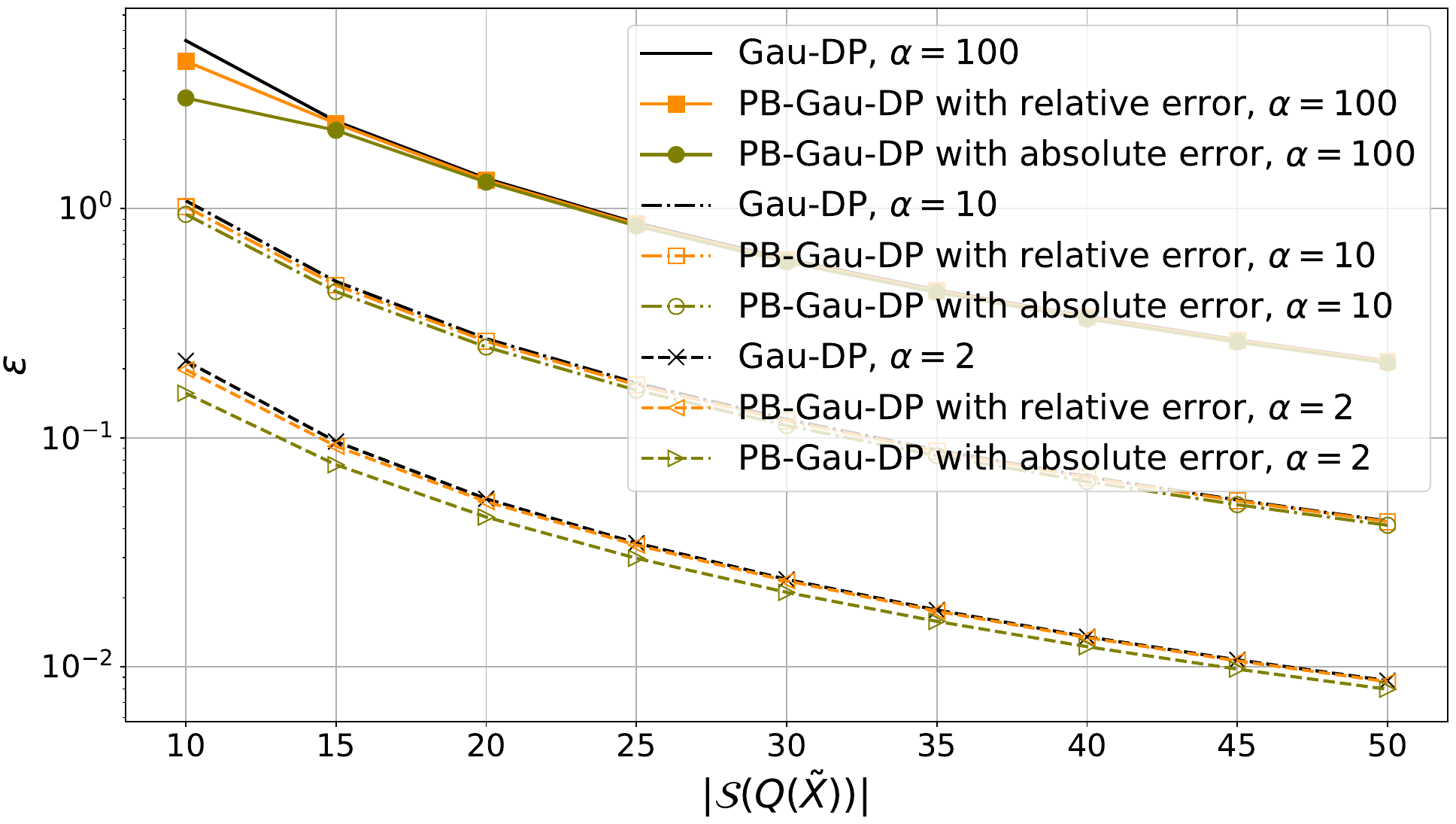}
\label{fig:RDP_privacy1}}
\subfigure[RDP {$\rho = 0.8, \Delta_Q = 4$.}]
{\includegraphics[width=0.49\textwidth]{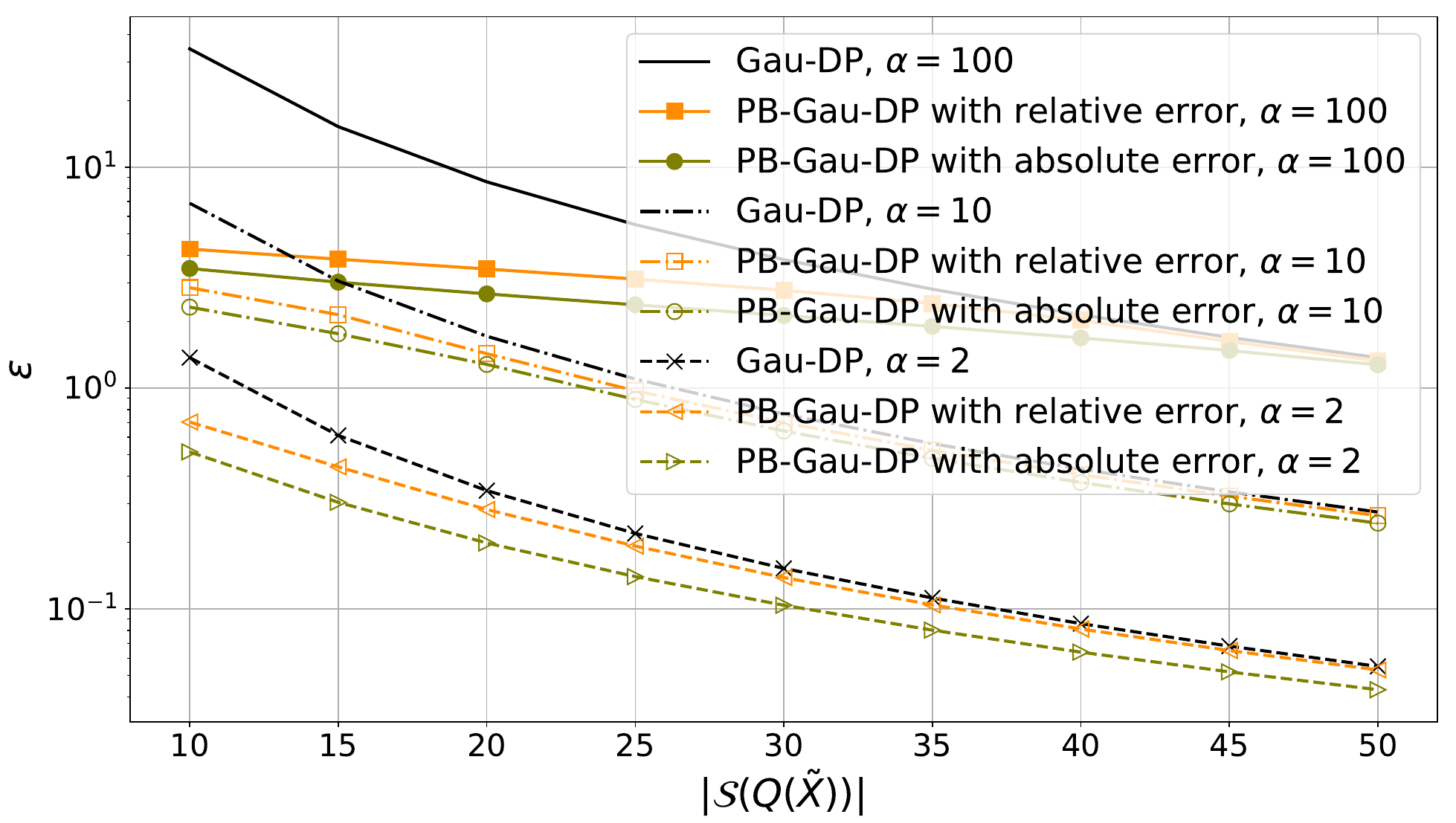}
\label{fig:RDP_privacy2}}
\caption{Boosted privacy comparison with fixed $\rho$, among mechanisms with relative error guarantee, mechanism with fixed output region, and mechanism with absolute error guarantee. Each mechanism shown in figure (a) and (b) achieves $(\epsilon,\delta)$-DP, and in figure (c) and (d) achieves $(\alpha,\epsilon)$-RDP. The preferred output regions for different mechanisms are aligned to be the same. Specifically, (a), (c) corresponds to a high sensitivity to aggregation ratio, and (b), (d) corresponds to low high sensitivity to aggregation ratio.}
\label{fig:privacy1}
\end{figure*}

\begin{figure*}[t]
\centering 
\subfigure[Approximate DP, {$\inf\mathcal{S}(Q(X)) = [-10, 10], \Delta_f = 2$.}]
{\includegraphics[width=0.49\textwidth]{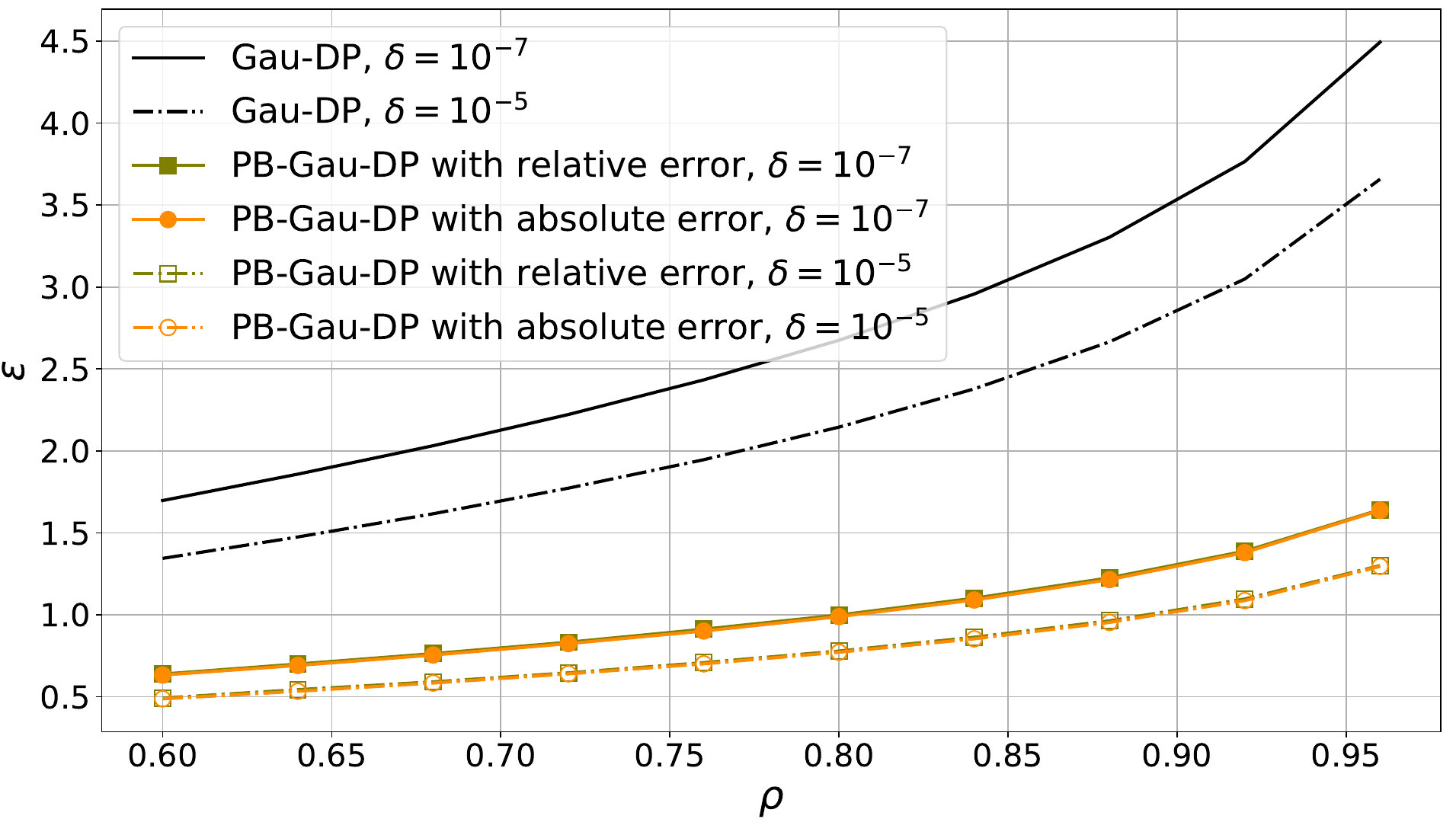}
\label{fig:DP_privacy21}}
\subfigure[Approximate DP, {$\inf\mathcal{S}(Q(X)) = [-10, 10], \Delta_f = 4$.}]
{\includegraphics[width=0.49\textwidth]{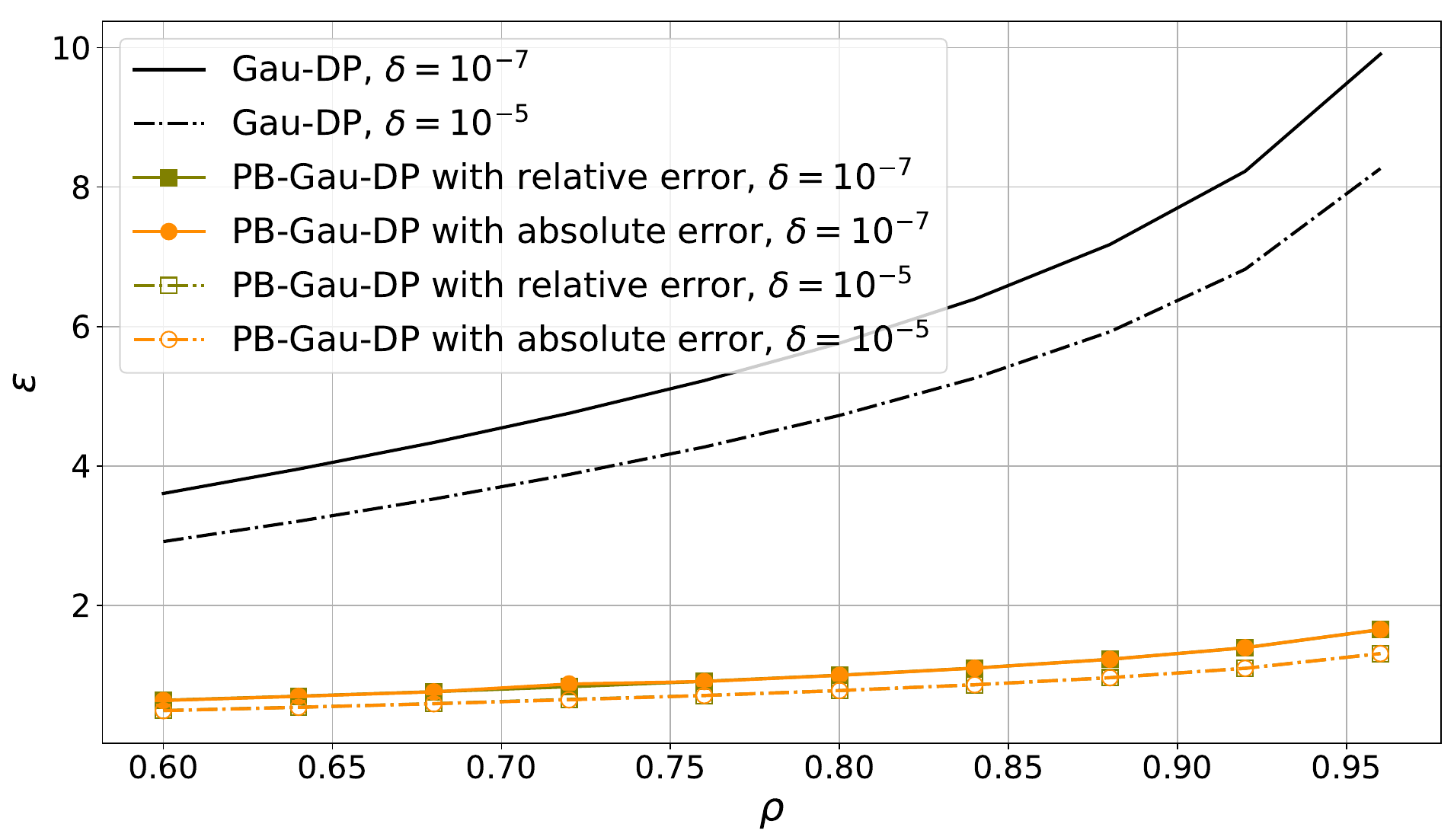}
\label{fig:DP_privacy22}}
\subfigure[RDP, {$\inf\mathcal{S}(Q(X)) = [-10, 10], \Delta_f = 1$.}]
{\includegraphics[width=0.49\textwidth]{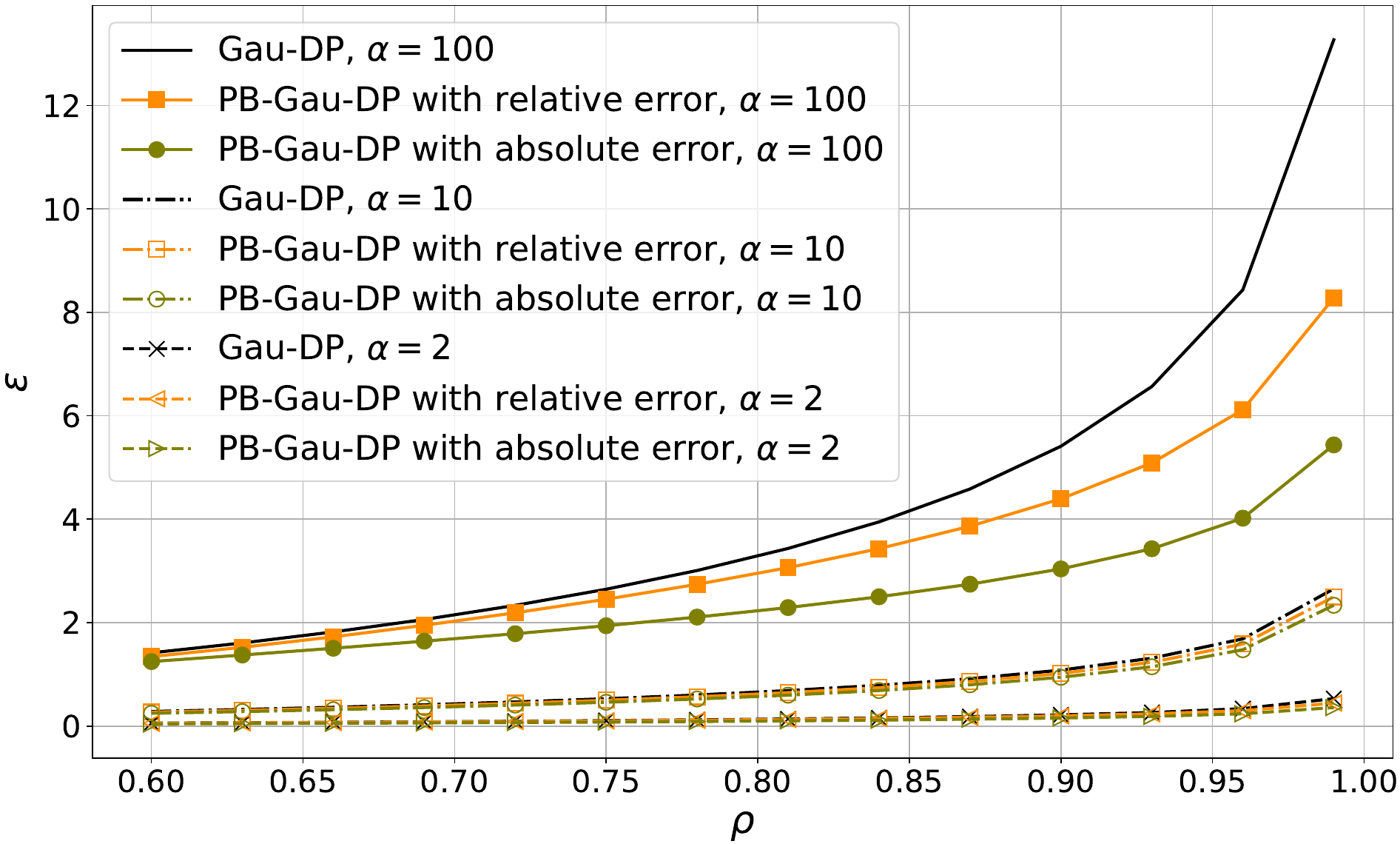}
\label{fig:RDP_privacy21}}
\subfigure[RDP, {$\inf\mathcal{S}(Q(X)) = [-10, 10], \Delta_f = 4$.}]
{\includegraphics[width=0.49\textwidth]{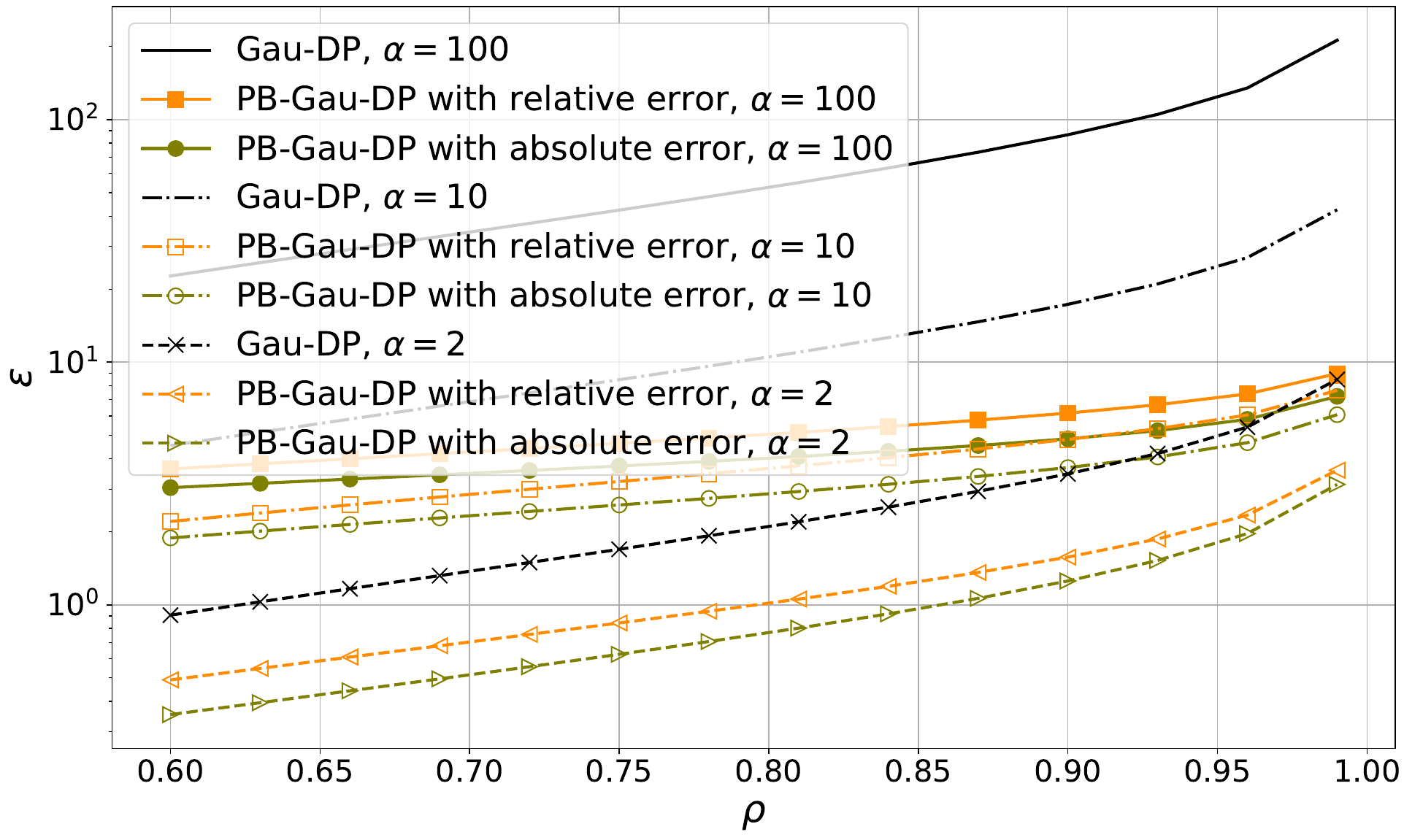}
\label{fig:RDP_privacy22}}
\caption{Boosted privacy comparison with fixed $\inf \mathcal{S}(Q(X))$, among mechanisms with relative error guarantee, mechanism with fixed output region, and mechanism with absolute error guarantee. Each mechanism shown in figure (a) and (b) achieves $(\epsilon,\delta)$-DP, and in figure (c) and (d) achieves $(\alpha,\epsilon)$-RDP. The preferred output regions for different mechanisms are aligned to be the same. Specifically, (a), (c) corresponds to a high sensitivity to aggregation ratio, and (b), (d) corresponds to low high sensitivity to aggregation ratio.}
\label{fig:privacy2}
\end{figure*}

\begin{figure}
    \centering
    \includegraphics[width=0.9\linewidth]{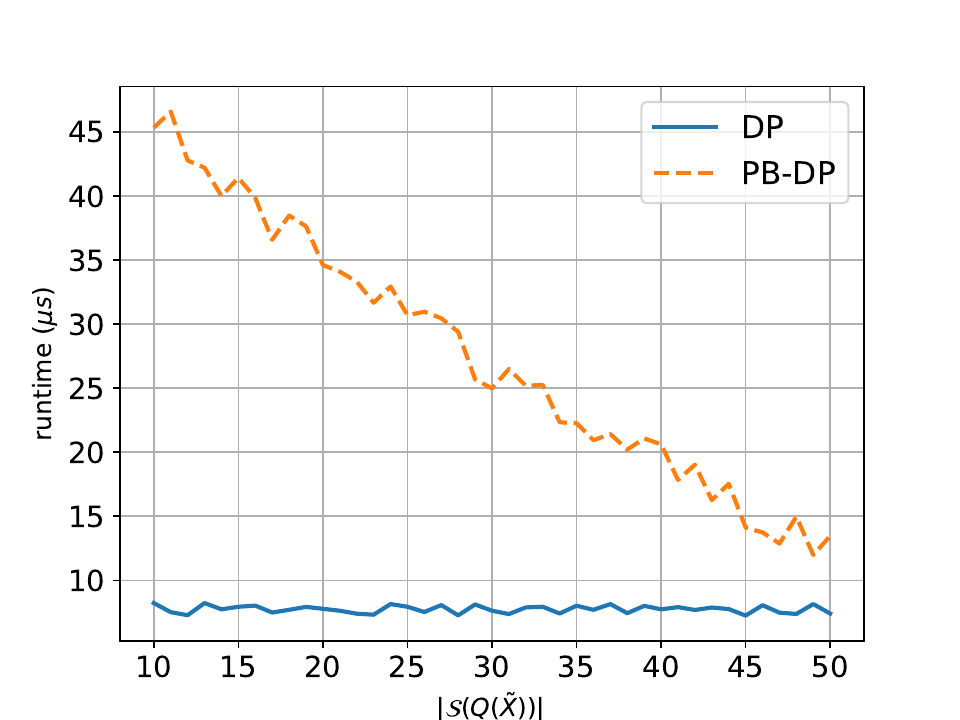}
    \caption{Runtime comparison between Gaussian-DP and PB-DP with Gaussian Kernel for different $|\mathcal{S}|$. }
    \label{fig:runtime}
\end{figure}

Recall that for general randomize response that achieves $\epsilon$-LDP, 
\begin{equation*}
    \begin{aligned}
        \begin{dcases}
            &p = \frac{e^{\epsilon}}{e^{\epsilon} + |\mathcal{X}| -1},\\
            &q = \frac{1}{e^{\epsilon} + |\mathcal{X}| -1},
        \end{dcases}
    \end{aligned}
\end{equation*}
where $p$ denotes the probability of direct release, and $q$ denotes the probability that the input data is perturbed to any other item. This is equivalent to the privacy boosting mechanism described in Proposition 6 when $\epsilon_0 = \epsilon$, which implies the mechanism grants no budget to the boosting rate $q$. On the other hand, when the mechanisms grant all budget to the boosting rate $q$, implying a zero $\epsilon_0$. The mechanism becomes the following:
\begin{equation*}
    \begin{aligned}
    \begin{dcases}
        &p = p_s = e^{\epsilon}/(|\mathcal{S}| e^{\epsilon} + |\mathcal{X}| - |\mathcal{S}|);\\
        &p_{\bar{s}} =  1/(|\mathcal{S}| e^{\epsilon} + |\mathcal{X}| - |\mathcal{S}|).\\
    \end{dcases}
    \end{aligned}
\end{equation*}


\subsubsection{Frequency Estimation Protocol}

Now consider there are $N$ users in the system, each holding a true value of $x_i$ for user index $i$. Each user locally privatizes his/her data with the privacy boosting LDP mechanism described above before submitting to the server. It is assumed that each $x_i$ belongs to its specific category. The server, after observing each user's submission, tries to aggregate the frequency of each category and each value. In the following, we use $F(\mathcal{S})$ to denote the true frequency of the appearance of any value belongs to category $\mathcal{S}$ and $F(x)$ as the true frequency of the appearance of a specific value $x$. Then the estimator for frequency estimation is shown as follows.

 \noindent\textbf{Frequency estimation for each category}:
The estimator for each category is 
\begin{equation}\label{est_s}
    \hat{F}_{\mathcal{S}} = \frac{\sum_{i=1}^N \mathbbm{1}_{\{x_i\in\mathcal{S}\}} - N |\mathcal{S}|p_{\bar{s}}}{p + (|\mathcal{S}| - 1)p_{s}-|\mathcal{S}|p_{\bar{s}}}.
\end{equation}
After obtaining an estimation on $\hat{F}_{\mathcal{S}}$, the server can further estimate the frequency of each element in this category.

 \noindent\textbf{Frequency estimation for each data value}:
From there, the estimator for each data can be obtained as
\begin{equation}\label{est_y}
    \hat{F}_{x} 
    =\frac{\sum_{i=1}^N  \mathbbm{1}_{\{x_i = x\}} - \hat{F}_{\mathcal{S}}(p_{s}-p_{\bar{s}})  - Np_{\bar{s}}}{p - p_s}.
\end{equation}


\begin{prop}
    The estimators in \eqref{est_s} and \eqref{est_y} are unbiased.
\end{prop}

With our privacy boosting mechanism, the data curator is able to estimate the frequency of each category and each item at the same time with high accuracy. Depending on the preference of the accuracy of $F_{\mathcal{S}}$ and $F_{x}$. our framework has the option to adjust $\epsilon_0$ to adjust the accuracy of $\hat{F}_{\mathcal{S}}$ and $\hat{F}_{x}$. We will provide more numerical analysis in Section. \ref{sec:exp-ldp}.

\section{Experiments}
In this section, we conduct a series of experiments to evaluate the performance and advantages of our proposed privacy-boosting differentially private (PB-DP) mechanisms. We start by comparing the privacy boosting capabilities of various mechanisms using the Gaussian mechanism as the kernel DP mechanism. We then explore the enlarged feasibility in the privacy profile for mechanisms with fixed preferred utility region, demonstrating how PB-DP can achieve smaller $(\epsilon, \delta)$ values that are not feasible with traditional bounded DP mechanisms. Next, we investigate the composed leakage comparison, showing the significant privacy improvements achieved through sequential composition of PB-DP mechanisms. Finally, we experiment with real data to illustrate the tradeoff between category frequency and item frequency using our PB-GRR mechanism on the \textit{Adult Dataset}.

\subsection{Privacy Boosting under Absolute Error and Relative Error Constraints}\label{sec:exp1}

In this experiment, we use the Gaussian mechanism as the kernel mechanism in our PB-DP framework and compare it against the Gaussian mechanism with privacy parameters that satisfy the same utility constraints. 
The overall privacy loss of our mechanism depends on several factors: a) the preferred region $\{S(y)\}_{y\in Y}$;  b) query sensitivity $\Delta_Q$, c) the confidence level $\rho$. We list our choice of these factors below.

We consider  two scenarios for query sensitivity, $\Delta_Q = 1$ and $\Delta_Q = 4$, representing low and high sensitivity, respectively. We evaluate our mechanism under two special cases discussed in the previous section: preferred region defined by absolute error and relative error. Throughout our experiments, we fix  $\delta$ for ($\epsilon$, $\delta$)-DP and $\alpha$ for RDP, and only compare the corresponding $\epsilon$. We perform the comparisons under two different choices of $\delta = 10^{-5}$ and $\delta = 10^{-7}$ for approximate DP, and three different choices of $\alpha = 2$, $\alpha = 10$ and $\alpha = 100$ for RDP.

We consider the output domain $\mathcal{R}=(-\infty, \infty)$. A critical factor in determining the strictness of utility constraints is the smallest size of the preferred region. Specifically, we define
$$\Tilde{X} = \arg \min_{X} p_{\mathcal{S}(Q(X))}.$$
Then the corresponding preferred region is $\mathcal{S}(Q(\Tilde{X}))$ with size
$|\mathcal{S}(Q(\Tilde{X}))|$. (Here, we slightly abuse the notation to also represent the size for continuous sets.)

To illustrate the privacy-utility tradeoff, we first fix $\rho=0.9$ for small sensitivity and $\rho=0.8$ for large sensitivity and vary the level of strictness of utility constraints measured by $|\mathcal{S}(Q(\Tilde{X}))|$, and plot the corresponding privacy loss $\epsilon$ in Figure \ref{fig:privacy1}.
We then fix the smallest preferred region $\mathcal{S}(Q(\Tilde{X}))$, and vary $\rho$ from $0.6$ to $1$ in Figure \ref{fig:privacy2}. In each figure, we demonstrate several settings: (a) small $\Delta_Q = 1$ with approximate DP; (b) large $\Delta_Q = 4$ with approximate DP; (c) small $\Delta_Q = 1$ with RDP; (d) large $\Delta_Q = 4$ with RDP.

In both Fig. \ref{fig:privacy1} and Fig. \ref{fig:privacy2}, we observe that using our PB-DP framework can reduce the required $\epsilon$ compared with the Gaussian mechanism under the same utility constraints, resulting in enhanced privacy. For low-sensitivity queries, our PB-DP mechanism consistently achieves lower $\epsilon$ values. The gap widens significantly for high-sensitivity settings ((b) and (d) compared to (a) and (c)). As the confidence level $\rho$ increases, the required $\epsilon$ generally increases for both mechanisms. However, the increment is less steep for our PB-DP mechanism, demonstrating its efficiency in maintaining lower privacy loss even under stringent utility constraints. Moreover, with more stringent DP requirements (small $\delta$ for approximate DP and large $\alpha$ for RDP), our PB-DP mechanism shows a more significant reduction in $\epsilon$. This demonstrates the efficiency of our mechanism in enhancing privacy without compromising utility.

\begin{figure*}[t]
\centering 
\subfigure[Privacy profile comparison between PB-DP and Bounded DP]
{\includegraphics[width=0.478\textwidth]{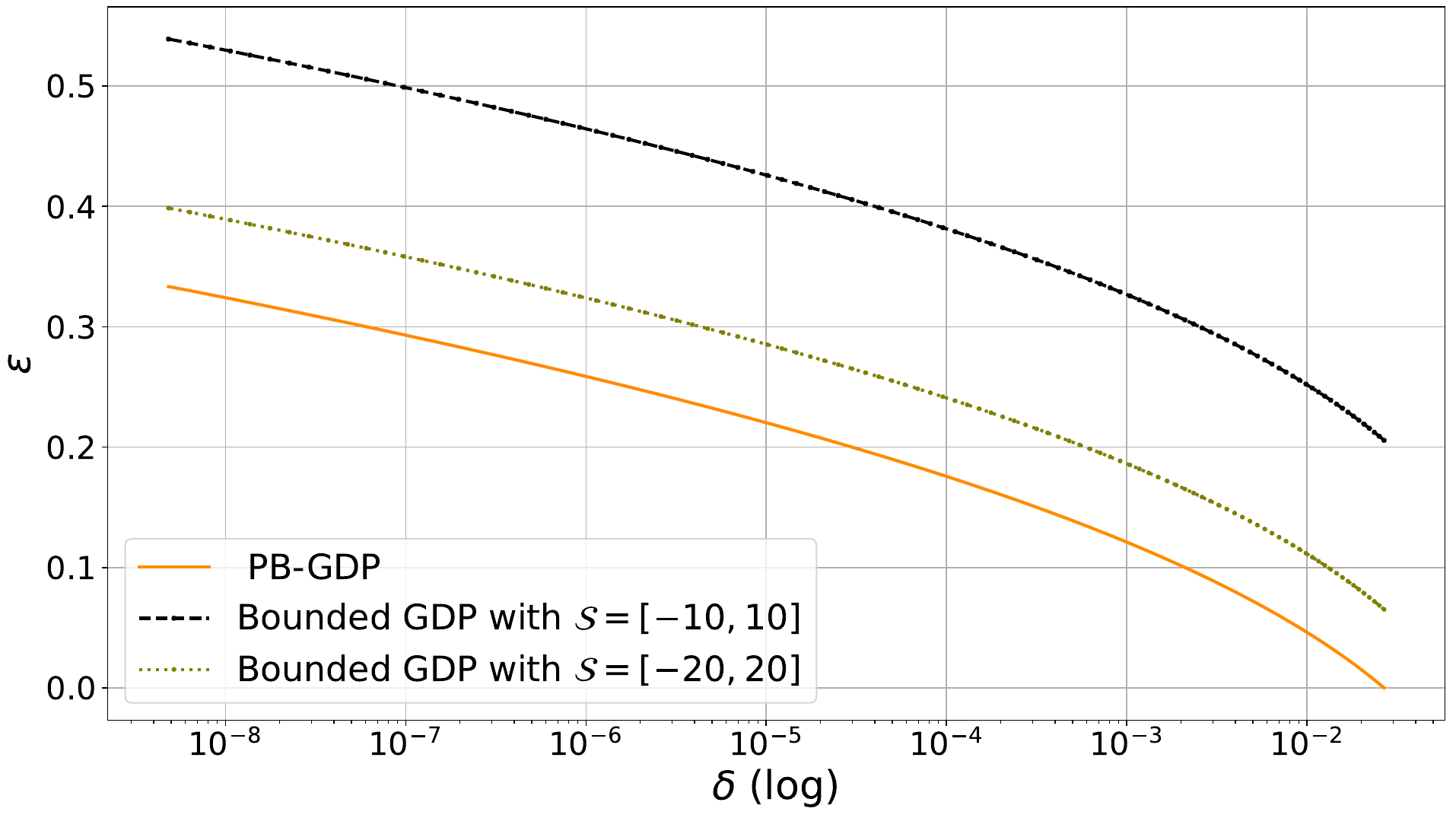}
\label{feasible_profile}}
\subfigure[Privacy boosting by reducing $\rho$.]
{\includegraphics[width=0.502\textwidth]{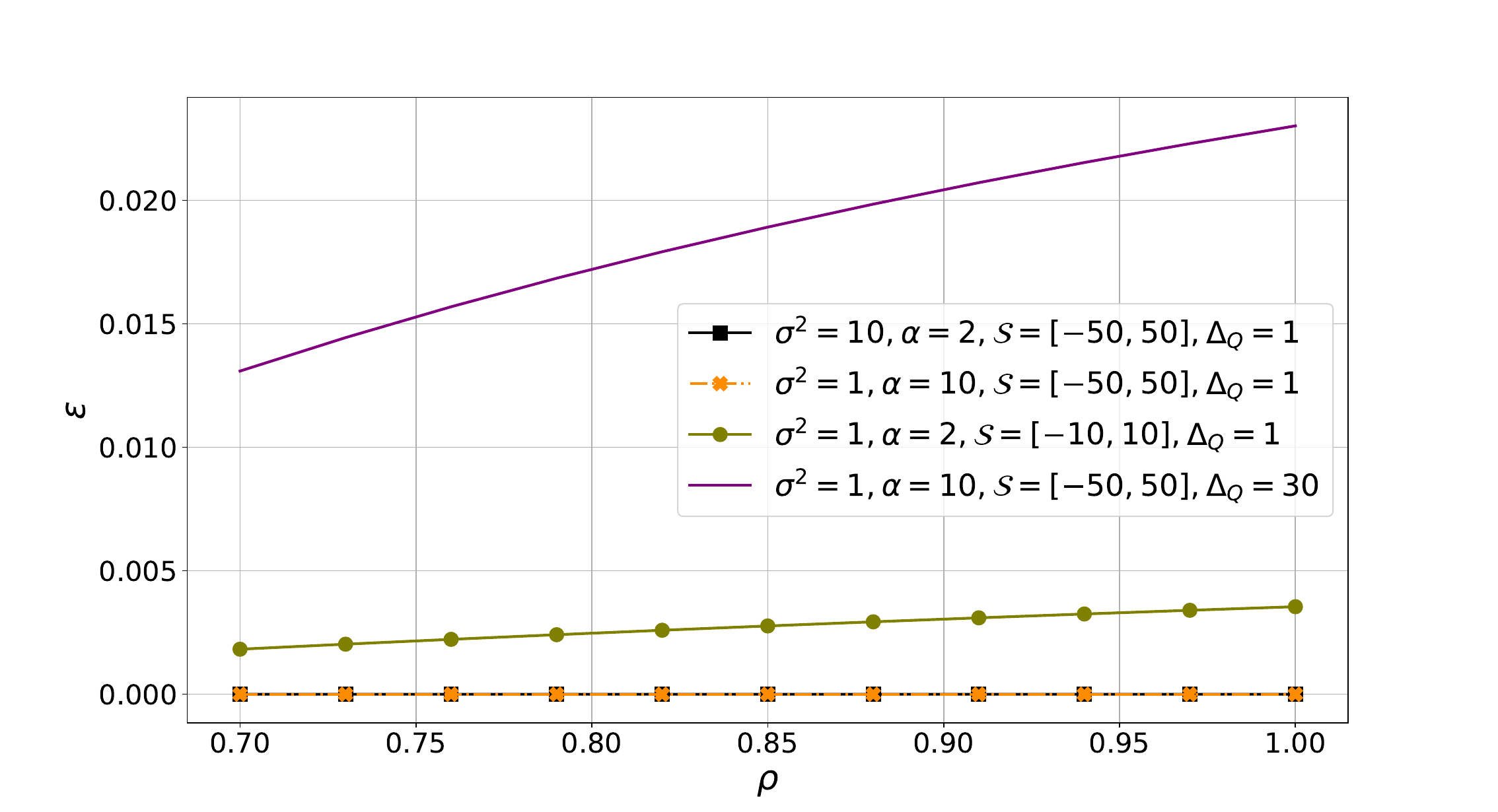}
\label{prob_ep}}
\caption{Comparison of the feasibility in the privacy profile for bounded Gaussian DP and PB-GDP with a fixed preferred region:
(a) Compares bounded GDP with PB-GDP for $\rho = 0.8$ and various $\mathcal{S}$.
(b) shows additional leakage caused by increasing $\rho$.}
\label{fig:profile}
\end{figure*}

\begin{figure*}[t]
\centering 
\subfigure[Post-composition leakage comparison with fixed likelihood of the preferred region.]
{\includegraphics[width=0.475\textwidth]{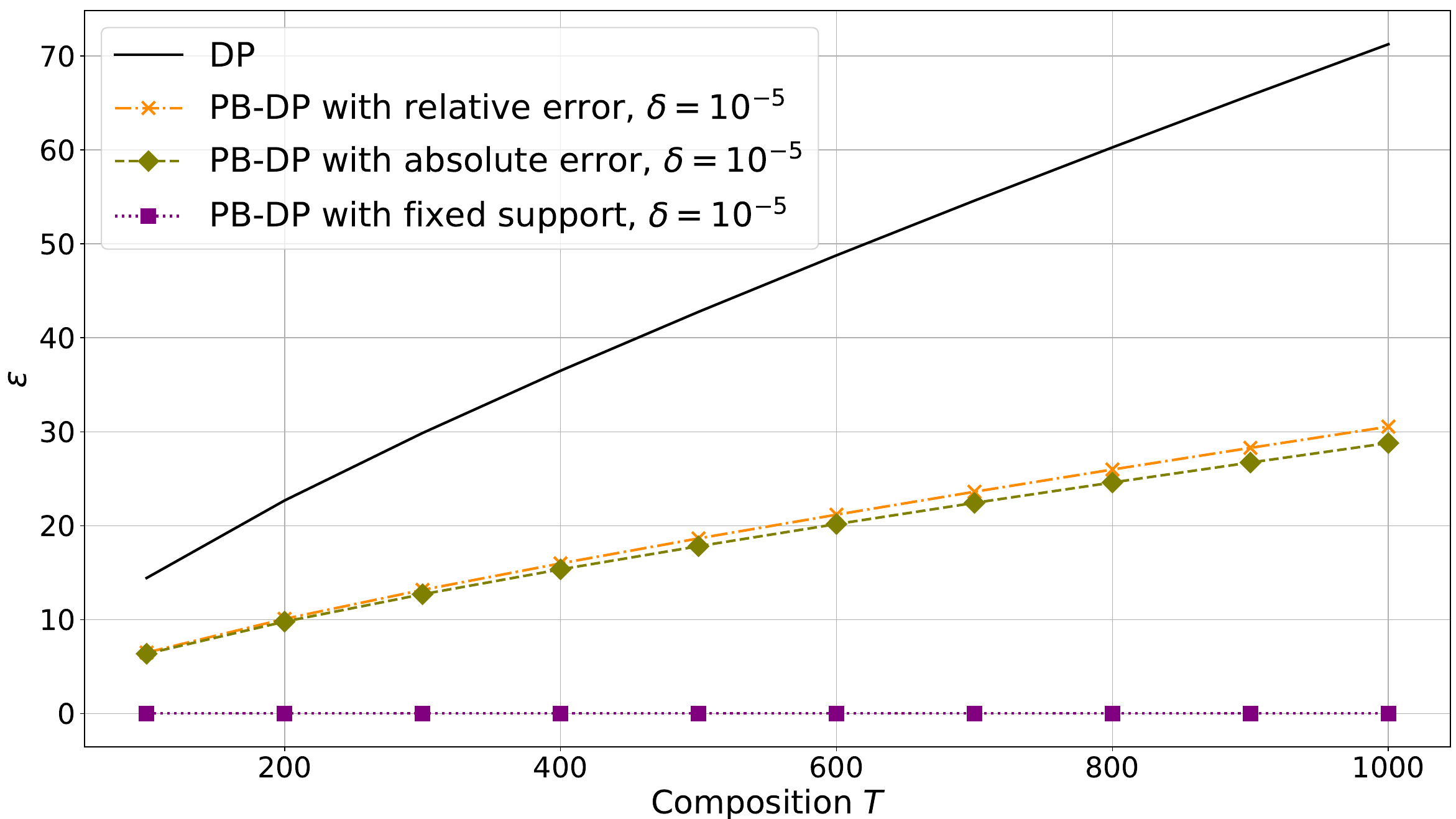}
\label{composition_dp}}
\subfigure[Post-composition leakage comparison with fixed privacy budget $\epsilon = 0.1$.]
{\includegraphics[width=0.49\textwidth]{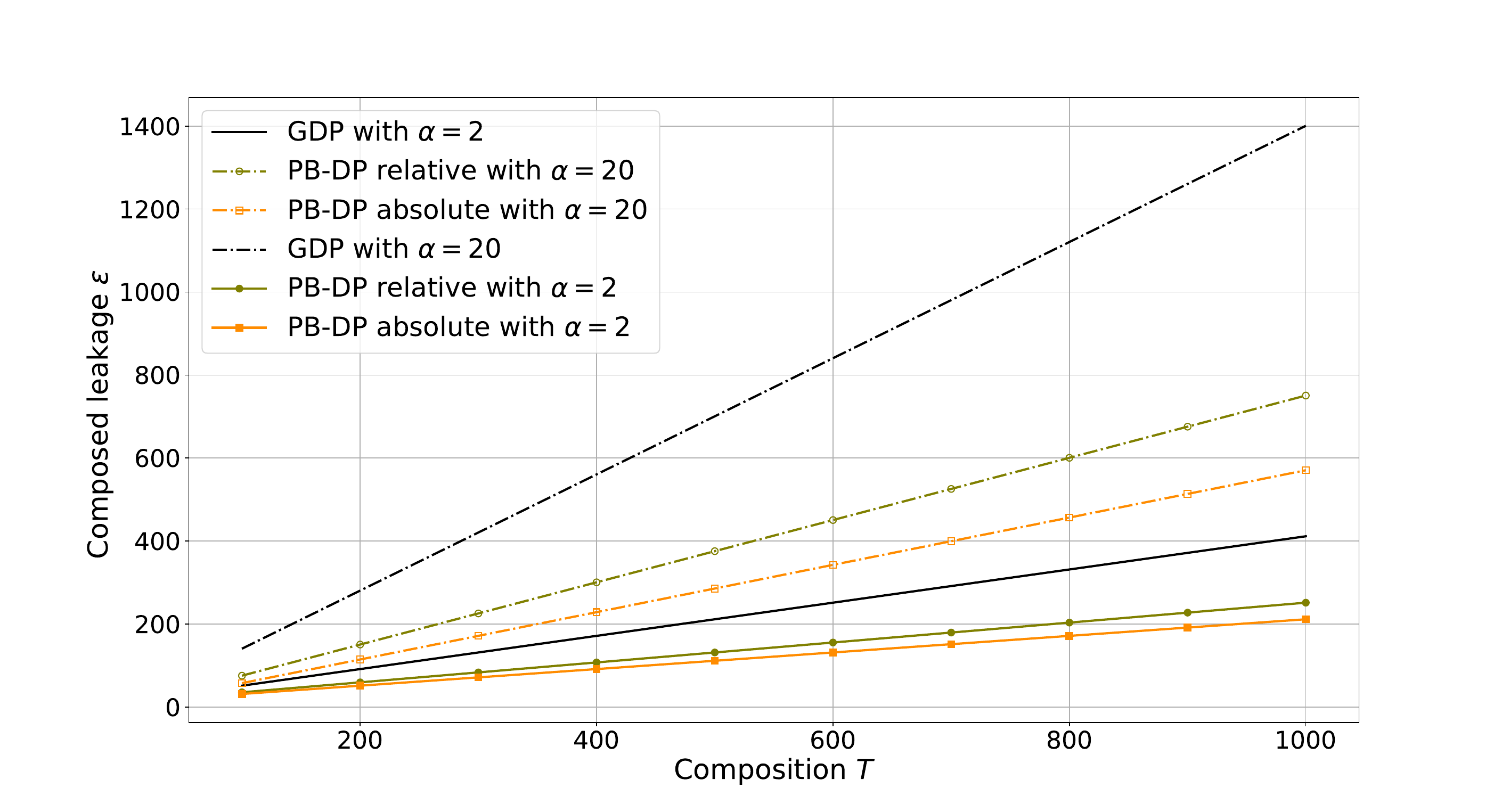}
\label{composition_rdp}}
\caption{Comparison of $T$-fold non-adaptive composed leakages among different mechanisms when each mechanism achieves a $\rho = 0.9$ with $\inf \mathcal{S}(Q(X)) = 10$, $\Delta_f = 3$. (a) compares post-composition leakage measured by $(\epsilon,\delta)$-DP, (b) compares post-composition measured by $(\alpha,\epsilon)$-RDP.}
\label{fig:composition}
\end{figure*}

\begin{figure*}[t]
\centering 
\subfigure[$|\mathcal{S}| = 10$, $\epsilon = 5$]
{\includegraphics[width=0.48\textwidth]{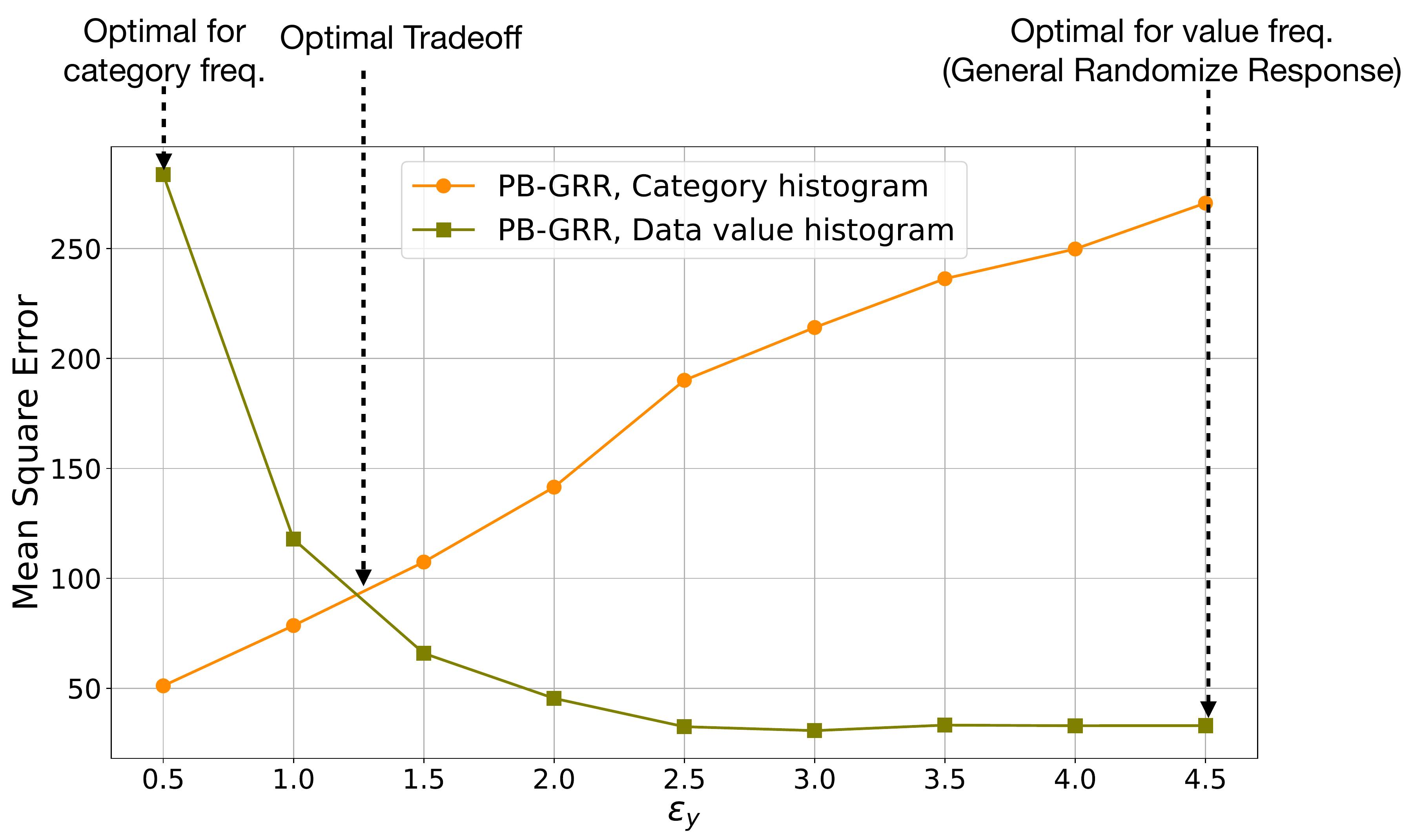}
\label{fig:pb-grr1}}
\subfigure[$|\mathcal{S}| = 5$, $\epsilon = 5$]
{\includegraphics[width=0.48\textwidth]{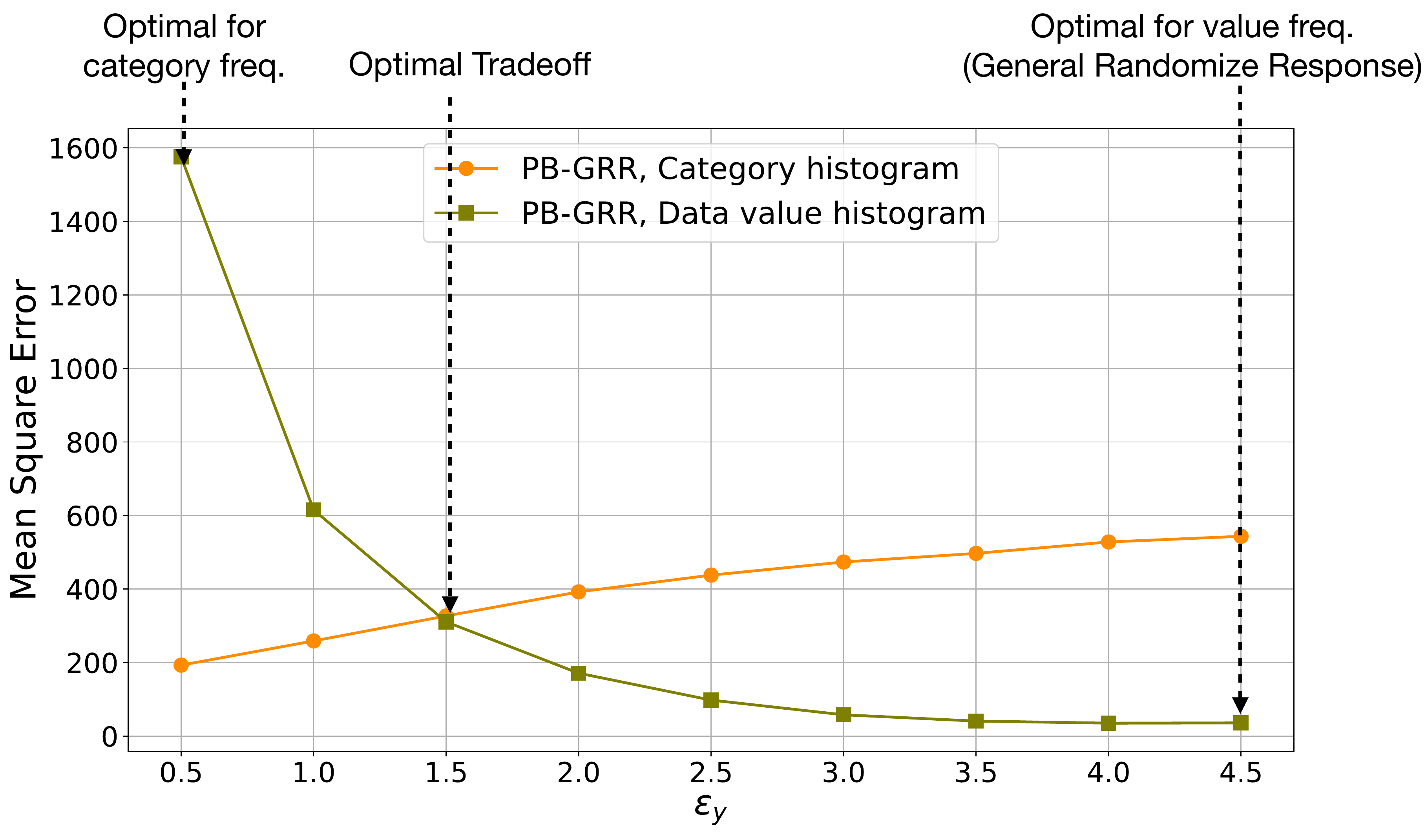}
\label{fig:pb-grr2}}
\caption{Accuracy comparison with real data ``Adult Income" between value frequency estimation and category frequency estimation with optimal $\epsilon_0$ for category frequency, value frequency, and optimal tradeoff. (a), (b) correspond to different sizes of $|\mathcal{S}|$.}
\label{fig:pb-grr}
\end{figure*}

\subsection{Computation Overhead}\label{sec.runtime}

Next, we perform a runtime comparison between the Gaussian DP and PB-DP mechanisms using a Gaussian kernel. For each mechanism, the dataset size $|\mathcal{S}(Q(\Tilde{X}))|$ is varied from 10 to 50, while $\rho = 0.9$ and $\Delta_Q = 1$ are kept constant. For each value of $|\mathcal{S}(Q(\Tilde{X}))|$, we derive the optimal parameters for the PB-DP mechanism using Algorithm 2. We then generate 10,000 samples across 100 iterations and calculate the average runtime for each.

The experiments were conducted on a desktop equipped with an Intel Core i9-14900KF processor and 64 GB of RAM. The results of the comparison are shown in Fig. \ref{fig:runtime}. We observe that the Gaussian DP mechanism maintains a constant runtime for noise generation across different values of $|\mathcal{S}(Q(\Tilde{X}))|$, while the PB-DP mechanism's runtime decreases as $|\mathcal{S}(Q(\Tilde{X}))|$ increases. Intuitively, for larger $|\mathcal{S}(Q(\Tilde{X}))|$, $\bar{p}_{\mathcal{S}(Q(X))}$ decreases, causing the mechanism to release a noisy output within the preferred utility region with fewer iterations.

\subsection{Enlarged Feasibility in Privacy Profile for a  Fixed Region}\label{sec.expfixed}

Next, we present another advantage of PB-DP with a fixed preferred region compared to bounded DP mechanisms with $q = 1$. We argue that by using the soft-bounded boosting factor $q \leq 1$, the feasible region in the privacy profile can be enlarged. This means that some small $(\epsilon, \delta)$ values that are not achievable by bounded DP mechanisms become feasible for our PB-DP mechanism.

We illustrate this idea from two perspectives:

(a) We design PB-DP with a fixed preferred region using an $(\epsilon, \delta)$-Gaussian mechanism as the kernel mechanism, where $\epsilon_0 = 0.1$ is fixed. We set the confidence level $\rho = 0.8$. We compare against a bounded Gaussian DP mechanism ($q = 1$) with the bound specified as $\mathcal{S}$. In Fig. \ref{feasible_profile}, we compare the feasible regions in terms of the privacy profile for PB-DP with a fixed preferred region and bounded Gaussian DP. The results indicate that PB-DP can improve the feasible region in the DP privacy profile.

(b) We then show how much utility, in terms of $\rho$, a PB-DP mechanism needs to sacrifice to reduce $\epsilon$. Unlike in case (a), we do not set a fixed $\epsilon_0$ in kernel mechanism; rather, we fix its variance $\sigma^2$ and then we vary $\rho$ from $0.7$ to $1$ and derive the corresponding privacy loss with $q$. By fixing $\sigma^2$, we ensure that the noise variance in the preferred region is consistent, isolating the effect of $q$ on additional leakage.
Recall that, measured by RDP, the total privacy loss 
$$\epsilon = \epsilon_0 + \mathcal{L}_1.$$ 
We then plot the additional loss $\mathcal{L}_1$ besides the kernel DP of each case under $\alpha = 2$ and $\alpha = 10$, for preferred regions $\mathcal{S} = [-10,10]$, $\mathcal{S} = [-50,50]$, and for sensitivities $\Delta_Q = 1$ and $\Delta_Q = 30$, respectively.
In Fig. \ref{prob_ep}, we observe a reduction in $\epsilon$ as $\rho$ decreases. Notably, the decrements do not depend on $\alpha$ in RDP due to the same $\sigma^2$ in the DP kernel. Additionally, a narrower preferred region ($\mathcal{S} = [-10,10]$) and higher sensitivity ($\Delta_Q = 30$) result in even more significant $\epsilon$ reduction.

\subsection{Homogeneous Composition}

In this section, we demonstrate that under composition, the privacy boosting can be even more significant.
In the experiment, we set each mechanism being composed to achieve $\rho = 0.9$ with $|\mathcal{S}(Q(\Tilde{X}))| = 10$ and $\Delta_Q = 3$. We vary the number of compositions $T$ from $1$ to $1000$. We consider both approximate DP and RDP. For approximate DP, we use the composition accounting algorithm described in Algorithm \ref{algo1} with the Analytic Fourier Accounting algorithm to capture the composition of the kernel mechanism. For RDP, the composition is naturally tight, and for each $\alpha$, the composed leakage is equivalent to $T$ times the leakage of a single mechanism. 

The experimental results are shown in Fig. \ref{fig:composition}, where (a) compares the composed privacy loss captured by $(\epsilon, \delta)$-DP with $\delta = 10^{-5}$; (b) compares the composed privacy loss captured by $(\alpha, \epsilon)$-RDP with $\alpha = 2$ and $\alpha = 20$. In each case, the composed privacy loss by the PB-DP mechanism is significantly smaller compared to the Gaussian mechanism that achieves the same utility constraint. 
From Fig. \ref{composition_rdp}, we observe that with stringent privacy guarantees, the decrement in composed privacy loss provided by PB-DP becomes more pronounced.

\subsection{Tradeoff between Category Frequency and Value Frequency}\label{sec:exp-ldp}

We next experiment with real data to illustrate the features of our PB-GRR mechanism that achieves $\epsilon$-LDP as introduced in Section. \ref{sec:pb-grr}. We adopt the \textit{Adult Dataset}\cite{misc_adult_2} from UCI, which contains census information with $45,222$ records and $15$ attributes. The attributes include both categorical ones, such as race, gender, and education level, as well as numerical ones, such as capital gain, capital loss, and weight. In this experiment, we assume each user adopts our PB-GRR mechanism to release their \textbf{age}, which is preprocessed as integers within the range $[10, 100]$. We then set $|\mathcal{S}| = 10$ and $|\mathcal{S}| = 5$ respectively. We fix $\epsilon = 5$ and vary $\epsilon_0$ from $0$ to $5$ to compare the Mean Squared Error (MSE) of the estimation of category frequency and value frequency. These two cases are plotted in Fig. \ref{fig:pb-grr1} and Fig.\ref{fig:pb-grr2} respectively. 

Note that there are other mechanisms that could be more optimal than the general randomized response, such as optimal unary encoding-LDP\cite{203872}, optimal local hash-LDP\cite{203872}, RAPPOR\cite{Rappor}, or Count Mean Sketch\cite{LPSApple}. However, the optimality of these mechanisms over the general randomized response is relevant for cases where the data cardinality is large or $\epsilon$ is small. According to \cite{203872}, the general randomized response is still optimal when $|\mathcal{X}|< 3 e^{\epsilon} + 2$.

In Fig.\ref{fig:pb-grr}, we mark the optimal values of $\epsilon_0$ to achieve the minimal mean squared error (MSE) for category frequency, value frequency, and the optimal tradeoff, respectively. We observe that the general randomized response mechanism is optimal for value frequency estimation but is the worst for category frequency estimation. This suboptimality becomes even more significant for large $|\mathcal{S}|$. Note that the $\epsilon_0$ for the optimal tradeoff we marked is not necessarily the optimal choice for a PB-GRR mechanism: depending on the preference for better value frequency or better category frequency, one can adjust $\epsilon_0$ for the specific goal.

\section{Discussions}

In this section, we explore several additional use-cases of the PB-DP framework.


\subsection{Convert to a Utility Boosting Framework} 

Our current framework can be understood as follows: given the utility constraints for a specific query, measured by $\mathcal{S}(Q(\cdot))$ and $\rho$, our mechanism can effectively derive a small $\epsilon$ that achieves $(\epsilon, \delta)$-DP for any $\delta$ or $(\alpha, \epsilon)$-RDP for any $\alpha$. Taking $\epsilon_0$ and $q$ as intermediate parameters, this procedure can also be simply viewed as a mapping function $\Lambda$ such that
$$\Lambda(\mathcal{S}(Q(\cdot)), \rho, \delta/ \alpha) \to \epsilon. $$
It is important to note that this mapping relationship can also work numerically in the opposite direction. To this end, we provide two additional interpretations of our framework.

 \noindent\textbf{Boosting $\rho$ for any given $\epsilon$:}
$$\Lambda^{-1}(\mathcal{S}(Q(\cdot)),  \epsilon, \delta/ \alpha) \to \rho, $$
This can be achieved by varying $\rho$ until the total leakage $\epsilon'$ is less than a desired value. Since $\rho$ monotonically increases with $\epsilon$, this functionality can be effectively achieved using a binary search. Boosting the confidence in a given interval for a DP mechanism is equally important in DP implementation. While utility-first DP mechanisms minimize leakage subject to utility constraints, privacy-first mechanisms tend to fix the privacy budget to manage risk.

 \noindent\textbf{Narrowed confidence interval:}
$$\Lambda^{-1}(\rho,  \epsilon, \delta/ \alpha) \to \mathcal{S}(Q(\cdot)). $$ 

Similar to boosting $\rho$, our framework can also be adapted to narrow $\mathcal{S}(Q(\cdot))$ given $\epsilon$ and $\rho$. Since $\mathcal{S}(Q(\cdot))$ monotonically decreases with $\epsilon$, this functionality can also be efficiently achieved via binary search. Narrowing the confidence interval is crucial as it enhances the precision of the query results, providing more accurate and reliable information while still maintaining the desired privacy guarantees. This is particularly important in applications where precise data analysis is critical for decision-making.

\subsection{Extensions and applications}
This paper considers several special cases as potential PB-DP applications, and in the experiments, we primarily focused on the Gaussian mechanism as the DP kernel. However, our framework can be extended to a variety of applications and support multiple noise distributions in the DP kernel.

For instance, PB-DP can work adaptively for multiple releases, either in an online or offline manner, to save budget or achieve high utility. One example is releasing data with meaningful ordering, such as in A/B testing. In such cases, one can design data-dependent PB-DP mechanisms with raw data for offline release. Conversely, data-independent PB-DP mechanisms can be designed using previous noisy releases in an online manner. PB-DP can also be incorporated with a variety of additive noise distributions, such as the Laplacian mechanism, exponential mechanism, binomial mechanism, etc., depending on the specific application.

However, it is important to note that PB-DP may not be the best option for pure DP mechanisms without relaxation. As PB-DP's PLD is more concentrated in the high privacy regime, but this comes at the cost of a longer tail. This inevitably increases the worst-case leakage captured by pure DP, making PB-DP less suitable for applications requiring strict privacy guarantees without any relaxation.


\section{Conclusions}

In this paper, we propose a general privacy boosting framework (PB-DP) with utility guarantees, which achieves $(\epsilon, \delta)$-DP or $(\alpha, \epsilon)$-RDP. We consider a general type of utility definition, captured by a preferred region and the confidence of the likelihood that a noisy generation falls within this region. In our design, the noise distribution leverages three elements: the DP kernel distribution, the form of the utility region, and a boosting factor. We then derive the privacy loss distribution (PLD) for our mechanism as a function of these elements. For a given confidence level, our framework adaptively searches for the optimal parameters determining these elements to achieve minimal total leakage. We studied four special cases regarding data-dependent and data-independent utility regions and mechanism settings, deriving closed-form parameters in the PLD expression for each case. Our numerical evaluations and experiments with real data show that our framework can effectively reduce privacy leakage compared to traditional DP mechanisms under given utility constraints across various scenarios. Notably, the advantage of our framework becomes even more significant for larger sensitivity to the aggregation ratio, addressing an outstanding challenge in the DP research community.

\bibliographystyle{IEEEtran}
\bibliography{ref}

\appendices



\section{Validation of the noise distriution}\label{app.validdist}
\begin{proof}
Next, we show the proposed distribution is valid:
Obviously,  $0\le(1-q)\le1$, $0\le1-\bar{p}_{\mathcal{S}(Q(X))}q\le 1$. On the other hand:
\begin{equation}
    \begin{aligned}
    &\int_{-\infty}^{\infty}f_{\mathcal{M}_{pb}}(y) d_{y}\\
    =&\int_{y\in \mathcal{S}(Q(X))}\frac{f_{\mathcal{M}}(y)}{1-\bar{p}_{\mathcal{S}(Q(X))}q}d_{y}\\
    &~~~~~~~~+\int_{y \notin \mathcal{S}(Q(X))}\frac{f_{\mathcal{M}}(y)(1-q)}{1-\bar{p}_{\mathcal{S}(Q(X))}q}d_{y}\\
    =&\frac{1}{1-\bar{p}_{\mathcal{S}(Q(X))}q}\int_{y \in \mathcal{S}(Q(X))}f_{\mathcal{M}}(y) d_{y}\\
    &~~~~~~~~+\frac{(1-q)}{1-\bar{p}_{\mathcal{S}(Q(X))}q}\int_{y \notin \mathcal{S}(Q(X))} f_{\mathcal{M}}(y) d_{y}\\
    =&\frac{(1 - \bar{p}_{\mathcal{S}(Q(X))})}{1-\bar{p}_{\mathcal{S}(Q(X))}q}+\frac{(1-q)\bar{p}_{\mathcal{S}(Q(X))}}{1-\bar{p}_{\mathcal{S}(Q(X))}q}\\
=&1.
\end{aligned}
\end{equation}
This implies that the PB-DP has a valid noise distribution.
\end{proof}

\section{PLD of a BR-DP mechanism}

\begin{proof}
The leakage of our DP mechanism can be expressed as:
\begin{equation*}
    \log\left\{\frac{\text{Pr}(\mathcal{M}(Q(X))=y)}{\text{Pr}(\mathcal{M}(Q(X'))=y)}\right\},
\end{equation*}
where $X$ and $X'$ are neighboring datasets that have at most one element different from each other. Without loss of generality, let $X$ and $X'$ satisfy the following condition:
\begin{equation*}
    p_{\mathcal{S}(Q(X'))} \ge p_{\mathcal{S}(Q(X))}.  
\end{equation*}
Since the noisy distribution is not continuous, we next bound the leakage through the following three cases:

\textbf{Case 1:} $y\in\mathcal{S}(Q(X))\cap\mathcal{S}(Q(X'))$:
\begin{equation*}
\begin{aligned}
\frac{\text{Pr}(\mathcal{M}_{pb}(Q(X))=y)}{\text{Pr}(\mathcal{M}_{pb}(Q(X'))=y)}
=\frac{f_{\mathcal{M}(X)}(y)}{f_{\mathcal{M}(X')}(y)} \frac{1-\bar{p}_{\mathcal{S}(Q(X'))}q}{1-\bar{p}_{\mathcal{S}(Q(X))}q},
\end{aligned}
\end{equation*}
and the probability of incurring this leakage is:
\begin{equation*}
\begin{aligned}
    \text{Pr}(y\in\mathcal{S}(Q(X))\cap\mathcal{S}(Q(X'))),\\
\end{aligned}
\end{equation*}
which corresponds to two cases:
\begin{equation*}
    \begin{aligned}
    \begin{dcases} 
        \int_{\tau'_l}^{\tau_u} f_{\mathcal{M}(X)}(y) d_{y}, & \text{if } Q(X') = Q(X) + \Delta_f; \\
        \int_{\tau_l}^{\tau'_u} f_{\mathcal{M}(X)}(y) d_{y}, & \text{if } Q(X) = Q(X') - \Delta_f; \\
    \end{dcases}
\end{aligned}
\end{equation*}
\textbf{Case 2:} $y\in\mathcal{S}(Q(X))/\mathcal{S}(Q(X))\cap\mathcal{S}(Q(X'))$:
\begin{equation*}
    \begin{aligned}
&\frac{\text{Pr}(\mathcal{M}_{pb}(Q(X))=y)}{\text{Pr}(\mathcal{M}_{pb}(Q(X'))=y)}\\
&=\frac{f_{\mathcal{M}(X)}(y)}{f_{\mathcal{M}(X')}(y)}\cdot\frac{1-\bar{p}_{\mathcal{S}(Q(X'))}q}{1-\bar{p}_{\mathcal{S}(Q(X))}q}\cdot{\frac{1}{1-q}}.\\
\end{aligned}
\end{equation*}
Then the probability of case 2 can be derived as,
\begin{equation*}
\begin{aligned}
W_1 =  \text{Pr}(y\in\mathcal{S}(Q(X))/\mathcal{S}(Q(X))\cap\mathcal{S}(Q(X')))
\end{aligned}
\end{equation*}
Which corresponds to two cases:
\begin{equation*}
    \begin{aligned}
    \begin{dcases} 
        \int_{\tau_l}^{\tau'_l} f_{\mathcal{M}(X)}(y) d_{y}, & \text{if } Q(X') = Q(X) + \Delta_f; \\
        \int_{\tau'_u}^{\tau_u} f_{\mathcal{M}(X)}(y) d_{y}, & \text{if } Q(X) = Q(X') - \Delta_f. \\
    \end{dcases}
\end{aligned}
\end{equation*}
\textbf{Case 3:} $y\in\mathcal{S}(Q(X'))/\mathcal{S}(Q(X))\cap\mathcal{S}(Q(X'))$:
\begin{equation*}
    \begin{aligned}
&\frac{\text{Pr}(\mathcal{M}_{pb}(Q(X))=y)}{\text{Pr}(\mathcal{M}_{pb}(Q(X'))=y)}\\
&=\frac{f_{\mathcal{M}(X)}(y)}{f_{\mathcal{M}(X')}(y)}\cdot\frac{1-\bar{p}_{\mathcal{S}(Q(X'))}q}{1-\bar{p}_{\mathcal{S}(Q(X))}q}\cdot{(1-q)}.\\
\end{aligned}
\end{equation*}
Probability for case 3:
\begin{equation*}
\begin{aligned}
    W_2 = &\text{Pr}(y\in\mathcal{S}(Q(X'))/\mathcal{S}(Q(X))\cap\mathcal{S}(Q(X'))),\\
\end{aligned}
\end{equation*}
Which corresponds to two cases:
\begin{equation*}
    \begin{aligned}
    \begin{dcases} 
        \int_{\tau_u}^{\tau'_u} f_{\mathcal{M}(X)}(y) d_{y}, & \text{if } Q(X') = Q(X) + \Delta_f;\\
        \int_{\tau'_l}^{\tau_l} f_{\mathcal{M}(X)}(y) d_{y}, & \text{if } Q(X) = Q(X') - \Delta_f. \\
    \end{dcases}
\end{aligned}
\end{equation*}
\textbf{Case 4}: $y\notin\mathcal{S}(Q(X))\cup\mathcal{S}(Q(X'))$:
\begin{equation*}
    \begin{aligned}
&\frac{\text{Pr}(\mathcal{M}_{pb}(Q(X))=y)}{\text{Pr}(\mathcal{M}_{pb}(Q(X'))=y)}\\
&=\frac{f_{\mathcal{M}(X)}(y)}{f_{\mathcal{M}(X')}(y)}\cdot\frac{1-\bar{p}_{\mathcal{S}(Q(X'))}q}{1-\bar{p}_{\mathcal{S}(Q(X))}q}.\\
\end{aligned}
\end{equation*}
For Case 4:
\begin{equation*}
\begin{aligned}
 &\text{Pr}(y\notin\mathcal{S}(Q(X))\cup\mathcal{S}(Q(X'))),\\
\end{aligned}
\end{equation*}
 corresponds to two cases:
 \begin{small}
\begin{equation*}
    \begin{aligned}
    \begin{dcases} 
        \int_{-\infty}^{\tau_l} f_{\mathcal{M}(X)}(y) d_{y} + \int_{\tau'_u}^{\infty} f_{\mathcal{M}(X)}(y) d_{y}, & \text{if } Q(X') = Q(X) + \Delta_f \\
        \int_{-\infty}^{\tau'_l} f_{\mathcal{M}(X)}(y) d_{y} + \int_{\tau_u}^{\infty} f_{\mathcal{M}(X)}(y) d_{y}, & \text{if } Q(X') = Q(X) - \Delta_f \\
    \end{dcases}
\end{aligned}
\end{equation*}
\end{small}
Note that Case 1 and Case 4 incur the same leakage, and their probabilities can be combined, which becomes $1-W_1-W_2$. 
Denote $\mathcal{L}_1 \overset{\Delta}{=} \log\left\{\frac{1-\bar{p}_{\mathcal{S}(Q(X'))}q}{1-\bar{p}_{\mathcal{S}(Q(X))}q}\right\}$, $\mathcal{L}_2 = -\log(1-q)$.

As $\mathcal{L}_2 = -\log(1-q) \ge 0$, in the worst case, $W_1$ picks the maximum between:
\begin{equation*}
    W_1 =  \max\left\{\int_{\tau_l}^{\tau'_l} f_{\mathcal{M}(X)}(y) d_{y},
        \int_{\tau'_u}^{\tau_u} f_{\mathcal{M}(X)}(y) d_{y}\right\}.
\end{equation*}
On the other hand, $-\mathcal{L}_2 < 0$, and in the worst-case, $W_2$ takes the minimum of:
\begin{equation*}
     W_2 = \min\left\{\int_{\tau_u}^{\tau'_u} f_{\mathcal{M}(X)}(y) d_{y},
        \int_{\tau'_l}^{\tau_l} f_{\mathcal{M}(X)}(y) d_{y}\right\}.
\end{equation*}

Define a shifted PLD of $f_Z(z)$:
\begin{equation*}
    f'_Z(z) \overset{\Delta}{=} f_Z(z - \mathcal{L}_1).
\end{equation*}
Then the Privacy Loss Distribution can be represented as:
\begin{small}
\begin{equation*}
\begin{aligned}
     f_{\Gamma}(\gamma) = W_1 f'_Z(\gamma - \mathcal{L}_2)+ W_2 f'_Z(\gamma + \mathcal{L}_2) +W_3f'_Z(\gamma).
\end{aligned}
\end{equation*}
\end{small}
This completes the proof.




\end{proof}

\section{Proof the privacy profile}
\begin{proof}
We derive the privacy profile of the PB-DP mechanism via the definition of DP profile. 
\begin{equation}\label{eq:delta'}
    \begin{aligned}
    \delta' \ge &\mathbb{E}_{\Gamma}[\max\{0,1-\exp(\epsilon-\gamma)\}]\\
    =&\int_{\epsilon}^{\infty}(1-\exp(\epsilon-\gamma))f_\Gamma(\gamma)d\gamma,\\
    \end{aligned}
\end{equation}
where
\begin{equation*}
\begin{aligned}
&\int_{\epsilon}^{\infty}(1-\exp(\epsilon-\gamma))f_\Gamma(\gamma)d\gamma\\
=&(1-W_1-W_2) \int_{\epsilon}^{\infty}(1-\exp(\epsilon-\gamma))f'_Z(\gamma)d\gamma\\
+&W_1 \int_{\epsilon}^{\infty}(1-\exp(\epsilon-\gamma))f'_Z(\gamma-\mathcal{L}_2)d\gamma\\
+&W_2 \int_{\epsilon}^{\infty}(1-\exp(\epsilon-\gamma))f'_Z(\gamma-\mathcal{L}_3)d\gamma\\
=&W_1 \delta'_Z(\gamma - \mathcal{L}_2)+ W_2\delta'_Z(\gamma + \mathcal{L}_2)+ W_3 \delta'_Z(\gamma).
\end{aligned}
\end{equation*}
where $\delta'_Z(\epsilon) \overset{\Delta}{=} \delta_Z(\epsilon-\mathcal{L}_1)$, denotes the shifted privacy profile of the kernel DP mechanism. This concludes the proof.
\end{proof}

\section{Proof of theorem 2}
\begin{proof}
Considering the four cases described in Appendix C. 

For case 1 and case 4:
\begin{equation*}
\begin{aligned}
    \log\left\{\frac{\text{Pr}(\mathcal{M}_{pb}(Q(X))=y)}{\text{Pr}(\mathcal{M}_{pb}(Q(X'))=y)}\right\}
    =&Z + \log\left\{\frac{1-\bar{p}_{\mathcal{S}(Q(X'))}q}{1-\bar{p}_{\mathcal{S}(Q(X))}q}\right\}\\
    \le &Z + \log\left\{\frac{1}{{1-q}}\right\},
\end{aligned}
\end{equation*}
For case 2:
\begin{equation*}
\begin{aligned}
    &\log\left\{\frac{\text{Pr}(\mathcal{M}_{pb}(Q(X))=y)}{\text{Pr}(\mathcal{M}_{pb}(Q(X'))=y)}\right\}\\
    =&Z + \log\left\{\frac{1-\bar{p}_{\mathcal{S}(Q(X'))}q}{1-\bar{p}_{\mathcal{S}(Q(X))}q}\right\} + \log\left\{\frac{1}{{1-q}}\right\}\\
    \le &Z + 2\log\left\{\frac{1}{{1-q}}\right\},
\end{aligned}
\end{equation*}
For case 3:
\begin{equation*}
\begin{aligned}
    &\log\left\{\frac{\text{Pr}(\mathcal{M}_{pb}(Q(X))=y)}{\text{Pr}(\mathcal{M}_{pb}(Q(X'))=y)}\right\}\\
    =&Z + \log\left\{\frac{1-\bar{p}_{\mathcal{S}(Q(X'))}q}{1-\bar{p}_{\mathcal{S}(Q(X))}q}\right\} + \log\left\{{{1-q_y}}\right\}\\
    \le &Z + \log\left\{{{1-q}}\right\},
\end{aligned}
\end{equation*}

Combine these three cases:
\begin{equation*}
\begin{aligned}
    &\log\left\{\frac{\text{Pr}(\mathcal{M}_{pb}(Q(X))=y)}{\text{Pr}(\mathcal{M}_{pb}(Q(X'))=y)}\right\}\le Z + 2\log\left\{\frac{1}{{1-q}}\right\}.
\end{aligned}
\end{equation*}

When the answer kernel DP mechanism satisfies $(\epsilon_0,\delta)$-DP, the following holds:
\begin{equation*}
    \text{Pr}\left\{Z\ge\epsilon_0\right\}\le\delta.
\end{equation*}
Then, 
\begin{equation*}
    \begin{aligned}
&\text{Pr}\left\{Z-2\log{(1-q)}\ge\epsilon_0-2\log{(1-q)}\right\}\le{\delta}.
\end{aligned}
\end{equation*}
which implies:
\begin{equation*}
    \text{Pr}\left\{\Gamma\ge\epsilon_0-2\log{(1-q)}\right\}\le{\delta}\\
\end{equation*}
To guarantee $(\epsilon,\delta)$-DP, $\epsilon\ge \epsilon_0 - 2\log{(1-q)}$
Therefore, the worst-case $q$ to guarantee $(\epsilon,\delta)$-DP for a given $(\epsilon_0,\delta)$-DP is $q = 1-e^{(\epsilon-\epsilon_0)/2}$.
\end{proof}

\section{Proof of Theorem 4}
\begin{proof}
By definition, the privacy loss distribution,
\begin{equation*}
\begin{aligned}
&{f}^2_{\Gamma}\left(\log\left(\frac{\text{Pr}(\mathcal{M}_0\cdot\mathcal{M}_1(X)=(y_0,y_1))}{\text{Pr}(\mathcal{M}_0\cdot\mathcal{M}_1(X')=(y_0,y_1))}\right)\right)\\
=&\text{Pr}(\mathcal{M}_0\cdot\mathcal{M}_1(X)=(y_0,y_1))
\end{aligned}
\end{equation*}
Due to the independence of $\mathcal{M}_0$ and $\mathcal{M}_1$
\begin{equation*}
\begin{aligned}
    &\text{Pr}(\mathcal{M}_0\cdot\mathcal{M}_1(X)=(y_0,y_1))\\=&\text{Pr}(\mathcal{M}_0(X)=(y_0))\text{Pr}(\mathcal{M}_1(X)=(y_1)),
\end{aligned}
\end{equation*}
on the other hand,
\begin{equation*}
\begin{aligned}
    &\text{Pr}(\mathcal{M}_0\cdot\mathcal{M}_1(X')=(y_0,y_1))\\=&\text{Pr}(\mathcal{M}_0(X')=(y_0))\text{Pr}(\mathcal{M}_1(X')=(y_1)),
\end{aligned}
\end{equation*}
Therefore, 
\begin{equation*}
\begin{aligned}
    &\log\left(\frac{\text{Pr}(\mathcal{M}_0\cdot\mathcal{M}_1(X)=(y_0,y_1))}{\text{Pr}(\mathcal{M}_0\cdot\mathcal{M}_1(X')=(y_0,y_1))}\right)\\
    =&\log\left(\frac{\text{Pr}(\mathcal{M}_0(X)=y_0)}{\text{Pr}(\mathcal{M}_0(X')=y_0)}\right)+\log\left(\frac{\text{Pr}(\mathcal{M}_1(X)=y_1)}{\text{Pr}(\mathcal{M}_1(X')=y_1)}\right)
\end{aligned}
\end{equation*}
and 
\begin{equation*}
\begin{aligned}
&{f}^2_{\Gamma}\left(\log\left(\frac{\text{Pr}(\mathcal{M}_0(X)=y_0)}{\text{Pr}(\mathcal{M}_0(X')=y_0)}\right)+\log\left(\frac{\text{Pr}(\mathcal{M}_1(X)=y_1)}{\text{Pr}(\mathcal{M}_1(X')=y_1)}\right)\right)\\
=&\text{Pr}(\mathcal{M}_0(X)=(y_0))\text{Pr}(\mathcal{M}_1(X)=(y_1)).
\end{aligned}
\end{equation*}
which implies that 
\begin{equation*}
\begin{aligned}
    {f}^2_{\Gamma}(\gamma)=&f_{\Gamma_0}(\gamma)\ast f_{\Gamma_1}(\gamma)\\
    =&f_{Z_0}(\gamma)\ast f_{R_0}(\gamma)\ast f_{Z_1}(\gamma)\ast f_{R_1}{(\gamma)}.
\end{aligned}
\end{equation*}
For independent and identical mechanisms,
\begin{equation*}
\begin{aligned}
{f}^T_{\Gamma}(\gamma)=&f_{\Gamma}(\gamma)\ast^T f_{\Gamma}(\gamma)\\
=&[(f'_{Z}\ast f_{R}) \ast^T (f'_{Z}\ast f_{R})](\gamma)\\
    =&[(f'_{Z}\ast^T f'_{Z}) \ast(f_{R}\ast^T f_{R})](\gamma),
\end{aligned}
\end{equation*}
where $\ast^T$ denote the operation of $T$-fold convolution, and
\begin{equation*}
\begin{aligned}
    &f_{R}\ast^T f_{R}(\gamma)\\
    = &\sum_{e_1+e_2\le T}\binom{T}{e_1,e_2} W_1^{e_1}W_2^{e_2}(1-W_1-W_2)^{T-e_1-e2}\\
    &\cdot \delta_{\text{Dirac}} (\epsilon - (e_1-e_2)\mathcal{L}_2)
\end{aligned}
\end{equation*}

This completes the proof.
\end{proof}

\section{Leakage and probabilities  for relative error}
\begin{proof}
Note that for positive $Q(X)$, 
\begin{equation*}
    \bar{p}_{\mathcal{S}(y)} = 1- \Phi_{\mathcal{M}}(\theta Q(X) + \tau) - \Phi_{\mathcal{M}}(-\theta Q(X) - \tau),
\end{equation*}
Then 
\begin{equation*}
\begin{aligned}
    \mathcal{L}_1 =& \max_{X, X'\in \mathcal{X}}\left\{\log\left(\frac{{1-\bar{p}_{\mathcal{S}(Q(X))}q}}{{1-\bar{p}_{\mathcal{S}(Q(X'))}q}}\right)\right\}\\
    &\le \log\left(\frac{{1-\bar{p}_{\mathcal{S}(0)}q}}{{1-\bar{p}_{\mathcal{S}(\Delta_f)}q}}\right),
\end{aligned}
\end{equation*}
on the other hand, 
The probability $W_1$ and $W_2$ can be specified as:

\begin{equation*}
    \begin{aligned}
    W_1 = &\max\int_{\min\{ \tau_u, \tau'_u \}}^{\max\{ \tau_u, \tau'_u \}}f_{\mathcal{M}(X)}(y) dy\\
=&\max\int_{\tau'_u}^{\tau_u}f_{\mathcal{M}(X)}(y) dy\\
=& \max(\Phi_{\mathcal{M}} (\tau_u') - \Phi_{\mathcal{M}}(\tau_u))\\
=& \Phi_{\mathcal{M}} (\Delta_Q\theta + \tau) - \Phi_{\mathcal{M}}(\tau)
    \end{aligned}
\end{equation*}
Similarly, $W_2 = \Phi_{\mathcal{M}}(\theta\Delta_Q - \tau) - \Phi_{\mathcal{M}}(-\tau)$.
 This concludes the proof.
\end{proof}




\section{Proof of parameters in fixed preferred region}

\begin{proof}
As $W_2=W_3=0$, the worst-case leakage exists when $\mathcal{L}_1$ achieves its maximum, where
$$\mathcal{L}_1 = \log\left\{\max\left\{\frac{{{p}_{\mathcal{S}(Q(X'))}}}{{p}_{\mathcal{S}(Q(X))}}, \frac{{{p}_{\mathcal{S}(Q(X))}}}{{p}_{\mathcal{S}(Q(X'))}}\right\}\right\}$$

which is maximized when $\frac{p_{\mathcal{S}(Q(X))}}{p_{\mathcal{S}(Q(X'))}}$ reaches its maximum. Therefore, the worst-case pair of $X$ and $X'$ can be obtained when:
\begin{equation*}
\begin{aligned}
    X, X' = &\argmax_{X,X'} \log\left\{ \frac{p_{\mathcal{S}(Q(X))}}{p_{\mathcal{S}(Q(X'))}}\right\}\\
    =&\argmax_{X,X'} \log(p_{\mathcal{S}(Q(X))}) - \log(p_{\mathcal{S}(Q(X'))})
\end{aligned}
\end{equation*}
This value is achieved when $Q(X) = \tau_l$, and $Q(X') = \tau_l + \Delta_f$.
This completes the proof.
\end{proof}

\section{Proof of Local Discrete Mechanism}
\begin{proof}
Recall that for general randomize response that achieves $\epsilon$-LDP, $p = \frac{e^{\epsilon}}{e^{\epsilon} + |\mathcal{Y}| -1}$ and $q = \frac{1}{e^{\epsilon} + |\mathcal{Y}| -1}$, where $p$ denotes the probability of direct release, and $q$ denotes the probability that the input data is perturbed to any other item.
From the expression of a UB-DP mechanism, to guarantee a pure $\epsilon$-LDP, $q = 1-e^{(\epsilon-\epsilon_0)}$. Then 
\begin{equation*}
    \begin{aligned}
    p=&\frac{ \frac{e^{\epsilon_0}}{e^{\epsilon_0} + |\mathcal{Y}| -1}}{1 - \frac{|\mathcal{Y}| - |\mathcal{S}|}{e^{\epsilon_0} + |\mathcal{Y}| -1} \frac{e^{\epsilon-\epsilon_0}-1}{e^{\epsilon-\epsilon_0}}}\\
    =& \frac{e^{\epsilon}}{e^{\epsilon} + (|\mathcal{S}| -1) e^{\epsilon-\epsilon_0} + |\mathcal{Y}| - |\mathcal{S}|},
    \end{aligned}
\end{equation*}
Similarly, 

\begin{equation*}
    \begin{aligned}
    p_{s}=&\frac{ \frac{1}{e^{\epsilon_0} + |\mathcal{Y}| -1}}{1 - \frac{|\mathcal{Y}| - |\mathcal{S}|}{e^{\epsilon_0} + |\mathcal{Y}| -1} \frac{e^{\epsilon-\epsilon_0}-1}{e^{\epsilon-\epsilon_0}}}\\
    =& \frac{e^{\epsilon-\epsilon_0}}{e^{\epsilon} + (|\mathcal{S}| -1) e^{\epsilon-\epsilon_0} + |\mathcal{Y}| - |\mathcal{S}|},
    \end{aligned}
\end{equation*}

and 

\begin{equation*}
    \begin{aligned}
    p_{\bar{s}}=&\frac{ \frac{1}{e^{\epsilon_0} + |\mathcal{Y}| -1}\frac{1}{e^{\epsilon-\epsilon_0}}}{1 - \frac{|\mathcal{Y}| - |\mathcal{S}|}{e^{\epsilon_0} + |\mathcal{Y}| -1} \frac{e^{\epsilon-\epsilon_0}-1}{e^{\epsilon-\epsilon_0}}}\\
    =& \frac{1}{e^{\epsilon} + (|\mathcal{S}| -1) e^{\epsilon-\epsilon_0} + |\mathcal{Y}| - |\mathcal{S}|},
    \end{aligned}
\end{equation*}
This completes the proof.
\end{proof}

\section{Unbiased estimator}
\begin{proof}
    The expectation of $\hat{f}_{\mathcal{S}}$ can be represented as:
    \begin{equation*}
    \begin{aligned}
        E[\hat{f}_{\mathcal{S}}] =& \frac{E\left[\sum_{i=1}^N \mathbbm{1}_{\{y_i\in\mathcal{S}\}}\right]- N |\mathcal{S}|p_{\bar{s}}}{p + (|\mathcal{S}| - 1)p_{s}-|\mathcal{S}|p_{\bar{s}}}\\
        =&\frac{f_{\mathcal{S}}(p+(|\mathcal{S}|-1)p_{s}) + (N-f_{\mathcal{S}}) |\mathcal{S}|p_{\bar{s}}- N |\mathcal{S}|p_{\bar{s}}}{p + (|\mathcal{S}| - 1)p_{s}-|\mathcal{S}|p_{\bar{s}}}\\
        =& f_{\mathcal{S}}.
    \end{aligned}
    \end{equation*}

On the other hand,
\begin{equation*}
\begin{aligned}
    E[\hat{f}_{y}] =&E[\frac{\sum_{i=1}^N  \mathbbm{1}_{\{y_i = y\}}] - E[\hat{f}_{\mathcal{S}}](p_{s}-p_{\bar{s}})  - Np_{\bar{s}}}{p - p_s}.\\ 
    =&\frac{f_{y} p - f_y p_s}{p - p_s} = f_y.
\end{aligned} 
\end{equation*}
This completes the proof.
\end{proof}

\newpage 

\appendices 

\section{Meta-Review}

The following meta-review was prepared by the program committee for the 2025
IEEE Symposium on Security and Privacy (S\&P) as part of the review process as
detailed in the call for papers.

\subsection{Summary}
The paper proposes a framework to improve the privacy of noise-adding DP mechanisms while respecting a given utility constraint on the query output. The proposed method reshapes the noise distribution with the goal of increasing the likelihood that query outputs under the DP mechanism fall within a given preferred region, based on a probability parameter.

\subsection{Scientific Contributions}
\begin{itemize}
\item Creates a New Tool to Enable Future Science
\item Addresses a Long-Known Issue
\item Provides a Valuable Step Forward in an Established Field
\end{itemize}

\subsection{Reasons for Acceptance}
\begin{enumerate}
\item The paper addresses the long-known issue of decreased query output utility under DP mechanisms. The proposed approach provides a creative solution to this issue by adapting the noise distribution based on desired constraints on the query output utility.
\item The proposed framework provides a significant step forward for the field. The authors' approach is technically novel and provides increased output utility compared to SOTA without the need for relaxation of the DP guarantees.
\end{enumerate}

\end{document}